\documentclass[letterpaper,12pt]{article}
\usepackage{tabularx} 
\usepackage{jheppub}
\usepackage{amsmath}  
\usepackage{graphicx} 
\usepackage{subcaption}

\begin{document}

\title{RG Flows and Thermofield-Double States in Holography}
\author{Suman Das, Arnab Kundu}
\affiliation{Theory Division, Saha Institute of Nuclear Physics, A CI of Homi Bhabha National Institute, 1/AF, Bidhannagar, Kolkata 700064, India.}
\emailAdd{suman.das[at]saha.ac.in, arnab.kundu[at]saha.ac.in}
\date{\today}

\abstract{
In this article, we consider a Renormalization Group flow of the Thermofield-Double state in a UV-complete description of Holography, by introducing a relevant deformation to the ${\cal N}=4$ super Yang-Mills theory at strong coupling. This RG-flow is known to have a non-trivial, interacting ${\cal N}=1$ fixed point at the IR. Geometrically, an RG-flow of the TFD-state naturally continues radially and inside the black hole event horizon and yields a Kasner-structure near singularity, as has been observed in recent works. We show that for a generic value of the deformation, the putative IR fixed point remains inside the black hole in a certain sense. By fine-tuning this deformation, the ``fixed point" can be brought arbitrarily close to the event horizon, while always remaining inside. Physically, this distinguishes between the vanishing temperature limit of the RG-flows with the one at exactly zero temperature. We further discuss its general implications in the context of Holography, including a study of the recently-proposed Holographic $a$-function, especially in the interior of the black hole. We also discuss how correlations between the two-copies of the TFD-state depend on this explicit RG-flow.
} 

\maketitle

\section{Introduction}

Renormalization Group (RG) flow is a foundational cornerstone in understanding Quantum Field Theories (QFT). Typically, given an ultra-violet (UV)-description (particularly with conformal symmetry) and an energy scale (associated to its relevant deformations), RG-flow determines how observables change as a function of this energy-scale. While this is usually performed in the vacuum-state of the theory, it is natural to consider an RG-flow of the excited states as well.

Thermal states are special. Even though the RG-evolution of a thermal state may not appear non-trivial, in the context of Holography, this problem is mapped to studying a general AdS-Schwarzschild black hole and its interior within the classical (super)gravity approximation. This idea was recently introduced in \cite{Frenkel:2020ysx}, further explored in {\it e.g.}~\cite{Hartnoll:2020rwq, Hartnoll:2020fhc}. This framework allows one to further sharpen aspects of quantum information for black holes in general, see {\it e.g.}~\cite{Caceres:2021fuw, Bhattacharya:2021nqj}.

Of particular interest is the thermo-field double (TFD)-state in gravity\cite{Israel:1976ur, Maldacena:2001kr} that represents the canonical purification of a thermal state. This is a particularly interesting state in the Hilbert space of the dual CFT, for several reasons. First, it demonstrates a clean version of the so-called canonical typicality\cite{2006}: by summing over either the left or the right degrees of freedom, the TFD-state density matrix yields an exact thermal density matrix. Secondly, recent progress in constructing traversable Wormholes in a UV-complete description of gravity is built on deforming a TFD-state, see {\it e.g.}~\cite{Gao:2016bin}. The TFD-state is definitely very important in distilling out the connections between quantum gravity and quantum information.

It is therefore interesting to explore how a Wilsonian RG-flow perspective affects such physical questions.\footnote{See {\it e.g.}~\cite{2021} for a recent study on coarse-graining the TFD state by using entanglement renormalization and a corresponding renormalization circuit.} On one hand, it is a natural set of questions to explore from the boundary dual CFT-perspective, especially so since the connection between an RG-flow perspective and the quantum information theoretic perspective is currently lacking. Although deep and important examples of such a connection are already understood, {\it e.g.}~between a monotonically decreasing $c$-function and (strong sub-additivity) inequalities of entanglement entropy\cite{Casini:2006es, Casini:2012ei}, including the presence of a boundary\cite{Casini:2016fgb}. On the other hand, in a Holographic framework, such an RG-flow is realized by considering a non-trivial back-reaction of a gravitational field propagating in the geometry. As alluded above, in \cite{Frenkel:2020ysx} it was demonstrated that such back-reactions will alter the classical inside of the black hole and enrich its structure. Armed with this, one can now pose well-defined physical questions addressing the quantum dynamics of information using these geometric backgrounds. For some early work along this direction, see {\it e.g.}~\cite{Albash:2011nq, Kundu:2020nir}.

In this article, we consider the effect of such an RG-flow when the relevant deformation at the UV can drives the system to an interacting IR CFT. For example, we explicitly consider the well-known example of mass deformation of the ${\cal N}=4$ super Yang-Mills (SYM) theory that leads to an interacting ${\cal N}=1$ fixed point\cite{Freedman:1999gp}, except now we consider an RG-flow of the TFD state in the ${\cal N}=4$ SYM induced by the deformation. This particular RG-flow places all related physics questions in arguably the best understood example of Holography. We also consider a simpler toy model example, which retains all salient qualitative physical features.

For the TFD-state, there are two natural scales in the system: Temperature $T$ of the mixed density matrix which is obtained by tracing out half of the degrees of freedom of the original TFD-state, and the coupling $\phi_0$ of the relevant deformation ${\cal O}$. Consider ${\rm dim}[{\cal O}] =\Delta$ for a $d$-dimensional CFT. The relevant parameter that controls the flow, from the dual CFT perspective, is $\phi_0/T^{d-\Delta}$. Geometrically, there are two radial scales: $r_0$ and $r_h$, where $r_0$ denotes the scale where back-reaction of the deformation becomes significant and $r_h$ denotes the location of the event horizon. In the gauge where the conformal boundary is located at $r \to 0$, $r_h \gg r_0$ corresponds to the ``low temperature" physics. In this case, one may hit the IR fixed point before reaching an IR horizon.\footnote{In the strict limit, this corresponds to the RG-flow of the vacuum state of the ${\cal N}=4$ SYM to the vacuum state of the ${\cal N}=1$ SYM theory. As we will momentarily argue, the zero temperature flow cannot be smoothly obtained from the TFD-state.} On the other hand, as we will explicitly show, when $r_h \sim r_0$, we still reach the limit $\phi_0/T^{d - \Delta} \to \infty$ because of the non-trivial scalar back-reaction.\footnote{Note that the statement of how close $r_0$ and $r_h$ are, is dependent on the radial gauge. As is well-known in AdS/CFT, the chosen radial coordinate, nonetheless, is physical. One can translate the comparison between the two radial scales into a comparison between two-point correlation functions in the corresponding geometric backgrounds. For our purposes, this is not needed.}

Curiously, we will observe that for a generic value of $\phi_0/T^{d-\Delta}$ the IR-fixed point of the (super)gravity potential lies inside the black hole event horizon, but always stays an infinite coordinate distance away from the singularity inside. By tuning $\phi_0/T^{d-\Delta} \to \infty$, this fixed point begins approaching the event horizon from inside but can never reach it. This is a geometric manifestation that one cannot access the zero temperature flow by tuning $\phi_0/T^{d-\Delta} \to \infty$ within the gravitational description.\footnote{Note, however, that if we consider an RG-flow from a UV CFT to an IR CFT in which the IR CFT still has relevant deformations turned on, the corresponding gravity dual, near the IR fixed point, will be described by an AdS to Kasner flow as in \cite{Frenkel:2020ysx}.} We will momentarily discuss a consequence of the classical statement on the properties of the spectrum of the boundary theory, under an RG-flow. We will further study simple probes of this RG-flow geometry, in terms of two-point correlation functions and entanglement entropy.

This article is divided into the following parts. In the next section we begin with a discussion on how a strict zero temperature physics cannot be obtained starting from a TFD state and within classical (super)gravity. We also discuss its simple ramifications on the spectrum of the theory. In section $3$, we discuss a simple toy model that can describe a Holographic RG flow from a UV CFT to an IR CFT. Subsequently, we discuss a supergravity model for the same in the next section. We review some basic features of the Wilsonian RG-flow of the corresponding coupling and offer some comments in section $5$. A discussion on the correlators and their behaviour under the RG-flow appears in the next section. We then explore the behaviour of the recently-proposed $a$-function associated to RG-flows of thermal states, within the models that we study here. Finally, we conclude with some future directions. Some technical details are relegated in several appendices.

\section{Some General Remarks on TFD-states}

We will primarily work with the TFD-state. This is given by
\begin{eqnarray}
\left|{\rm TFD} \right\rangle = \frac{1}{\sqrt{Z\left[\beta \right]}} \sum_{n=0}^\infty e^{-\beta E_n/2} \left| n \right\rangle_{\rm L} \otimes \left| n \right\rangle_{\rm R} \ ,\label{tfd}
\end{eqnarray}
where $Z[\beta]$ is a normalization which we assume remains finite for the entire range of $\beta$. As is well-known, the TFD-state is maximally entangled and tracing over one copy of the system ({\it e.g.}~over the Hilbert space spanned by $\{\left| n \right\rangle_{\rm L} \}$), one obtains a thermal density matrix, with a temperature $\beta^{-1}$. Given the TFD-state for a quantum system with a spectrum that is bounded from below, a zero temperature limit can be meaningfully taken, provided $\beta\to \infty$, $E_0\to 0$, such that $\beta E_0 \to 0$. Here $E_0$ is the energy of the ground state. In this limit, one simply obtains: $\left|{\rm TFD} \right\rangle = \left| 0 \right\rangle_{\rm L} \otimes \left| 0 \right\rangle_{\rm R}$, which is an unentangled state.\footnote{Note that, in case we have a non-vanishing zero point energy, $\beta E_0 \to \infty$ in the limit $\beta \to \infty$. However, the factor of $e^{-\beta E_0/2}$ is cancelled by the same term appearing in the denominator of (\ref{tfd}), due to the normalization. This yields the same result.}

Strictly speaking, the TFD-state becomes an unentangled product of the vacuum states, provided $\beta E_0 \to 0$ and $\beta E_i \to \infty$, where $E_i$ are the energies of the excited states. The latter is easily ensured by scaling $\beta \Delta E \to \infty$, where $\Delta E$ is the mass gap (energy difference between first excited state and the vacuum state).\footnote{We can also fix $\beta E_0= {\rm constant}$, which does not change the conclusion.} It is, in fact, sufficient to arrange appropriately the hierarchy of temperature with the mass gap in the spectrum. In the limit $\beta \Delta E \gg 1$, however, the state remains entangled: There are exponentially small but non-vanishing contributions from excited states in the spectrum. Physically, the difference arises from the difference between the $T\ll 1$ physics to the physics at $T \to 0$. In one case, there is small but non-vanishing entanglement (therefore partial tracing yields a thermal state, with a small temperature); in the other, entanglement vanishes (partial tracing yields a pure density matrix).

It is therefore already clear that a vanishing temperature limit of the TFD-state is subtle, for a generic quantum mechanical system. Consider now, a quantum mechanical system with a Hamiltonian $H_1$ whose spectrum is explicitly known. In particular, let us assume that it allows for a mass gap $\Delta E$ such that $\beta \Delta E \to \infty$ can be achieved. We can construct a TFD-state following the definition in (\ref{tfd}). Now consider introducing a deformation to this Hamiltonian, such that the new Hamiltonian is given by $H_{\rm new} = H_1 + H_2$, where $H_2$ is not necessarily a small perturbation. The spectrum will change and the new mass gap $\Delta E_{\rm new}$ will also allow a limit $\beta \Delta E_{\rm new} \to \infty$, unless $\Delta E_{\rm new} \to 0$ such that $\beta \Delta E_{\rm new}$ remains fixed at some value.

To summarize, it is therefore clear that a quantum mechanical system with a non-vanishing mass gap does naturally allows for a limit in which the TFD state becomes an unentangled product state of left and right vacua. For a spectrum with a vanishing mass gap, however, the TFD state remains entangled, even in the vanishing temperature limit. Given a Hamiltonian with a non-vanishing mass gap, it is certainly possible to introduce new interaction terms such that, in some parametric regime, the mass gap closes off (see {\it e.g.}~\cite{Anand:2017yij}). In such a case, although the TFD-state of the original Hamiltonian allows for a $T\to 0$ limit where it becomes unentangled, the TFD-state of the final Hamiltonian remains entangled. Vice-versa.

In this article, we consider CFTs (and its deformation) with a Holographic dual. In particular, we consider the TFD-state of the dual CFT. For CFTs, mass gap vanishes since conformal symmetry is incompatible with it. Geometrically, the TFD-state is represented by an eternal black hole geometry, with two asymptotic boundaries where two copies of the dual CFT are defined\cite{Israel:1976ur, Maldacena:2001kr}. Once the eternal black hole is constructed, there is no limit in which it will split into two disconnected empty AdS-geometries.\footnote{Such a process is not viable within Einstein-gravity as it involves a change in the topology. This would have to be mediated by some $1/N$-effect in the CFT, and therefore involves stringy-physics in the bulk.} In other words, once an event horizon is introduced, it will always yield a  non-trivial correlation functions between operators on the left and the right CFTs. See {\it e.g.}~\cite{Leutheusser:2021qhd} for a recent discussion on how such causal structures can arise from the operator algebra structure in the dual CFT.

In light of our previous discussion, the low temperature limit therefore corresponds to $\beta \Delta E = {\rm fixed}$ in these cases. We will moreover construct explicit TFD-states when the CFT is subject to a relevant deformation. Thus, we must also have, in the low temperature limit: $\beta \Delta E_{\rm new} = {\rm fixed}$. In turn we obtain $\lim_{\beta \to \infty} \Delta E / \Delta E_{\rm new} = {\rm constant}$. On the other hand, suppose $\phi_0$ corresponds to the coupling constant of the relevant deformation, we would expect a following relation:
\begin{eqnarray}
\frac{\Delta E_{\rm new}}{\Delta E} = \Phi\left(\frac{\phi_0}{T^{d-\Delta}} \right) \ ,
\end{eqnarray}
where $\Delta$ is the conformal dimension of the relevant operator, $d$ is the spacetime dimension of the CFT and $\Phi$ is an unknown function. The low temperature constraint above sets a mild constraint on this function: For example, $\Phi \sim \phi_0/T^{d-\Delta}$, as $T \to 0$ is not an allowed function. This clearly is a feature of large-$N$ theories with a Holographic dual description.

Note further that, a vanishing mass gap essentially implies that the spectrum does not exhibit level-repulsion, since at least two energy levels can come very close to each other. Level-repulsion is a tell-tale sign of quantum chaotic dynamics, and a late-time behaviour of the corresponding system is expected to be described by a random matrix theory: in particular the universal features of a dip-ramp-plateau behaviour of the spectral form factor. From our discussion above, it is evident that a semi-classical description based on the eternal black hole geometry does not exhibit level-repulsion. One requires a completely quantum gravitational description to capture them\cite{Cotler:2016fpe}. Such a description clearly includes $(1/N)$-effects and therefore is capable of capturing the limit in which the TFD-state becomes an unentangled product state of two vacuua.

We will now consider a TFD-state in a Holographic CFT$_d$, subject to a relevant deformation of dimension $\Delta$, closely following the lines of \cite{frenkel2020holographic}. We will remain within the classical geometrical description and hence the TFD-state always remains entangled in our description. However, since we introduce a dimensionful relevant coupling, it provides us with a natural vanishing temperature limit: $\phi_0/T^{d-\Delta} \to \infty$, while $\beta \Delta E$ remains fixed. We will consider a potential that allows for an UV CFT that flows to an IR CFT, due to the relevant deformation. There are standard RG-flow geometries that connect these fixed points at $T=0$. They describe a geometry that interpolates between two AdS-vacua, correspondingly. For the TFD-state, we will observe a richer physics which we discuss now.

\section{A Toy Model}

Let us begin with a toy model with a single scalar field in the bulk with the following bulk action:
\begin{equation}\label{action1d}
    \mathcal{S}=\int d^{d+1}x \sqrt{\mid{g}\mid}\left(\frac{1}{2\kappa^2}R-\frac{1}{2}g_{\mu\nu }\partial_{\mu}\phi \partial_{\nu}\phi-V(\phi)\right) \ . 
\end{equation}
We will momentarily focus on specific values of $d$, but our discussions will be applicable for general values. For concreteness, let us choose: $\kappa^2=1$ and $V(\phi)=\phi^3-\phi^2-3$. The potential is chosen such that there are two extrema: 
\begin{eqnarray}
\frac{\partial V}{\partial\phi} = 0 \quad \implies \phi = 0 , \frac{2}{3} \ .
\end{eqnarray}
Correspondingly, the values of the potential at these extrema are: $V(0)=-3$ and $V(2/3)= - 85/27 < V(0)$. Here, $\phi=0$ is the maximum and $\phi=2/3$ is the minimum of the potential. Correspondingly, there will be two AdS-solutions: one at $\phi=0$, which we call AdS$_{\rm UV}$ and one at $\phi=2/3$, which we refer to as AdS$_{\rm IR}$. We have further chosen a normalization for the curvature-scale of the AdS$_{\rm UV}$-geometry.

Given the above structure, within the solution space of Einstein-equations, there is a flow geometry that interpolates between the AdS$_{\rm UV}$, at the UV, and the AdS$_{\rm IR}$, at the IR. This corresponds to, in the dual gauge theory, an RG-flow from a UV fixed point to an IR fixed point. There are numerous interesting and rich classes of such flow examples in the literature, see {\it e.g.}~\cite{Kiritsis:2016kog}. Here, we will focus on the flow of a state instead, in particular of a thermal state, following the recent works in \cite{frenkel2020holographic}. A CFT thermal state is realized as a Black Hole solution in the geometry, and in the Euclidean description, one caps off the geometry at the corresponding event horizon. Therefore, any putative flow must stop at the horizon, which always plays the role of the IR.

In the Lorentzian description, however, event horizon is not a special point\footnote{We are assuming that there is {\it no drama} at the event horizon. At any rate, classically, nothing breaks down at this location.}, and one is able to continue the geometry inside the Black Hole. The standard example is, of course, that of the Schwarzschild geometry, and its subsequent Kruskal-extension. In case of an RG-flow geometry, the boundary CFT is perturbed by a relevant deformation that triggers the flow. In the dual geometric description this is captured by turning on {\it e.g.}~a scalar field in the bulk, which back-reacts and sources a non-trivial geometry. In a Lorentzian framework, the effect of this back-reaction can be computed explicitly, even within the event horizon and all the way to the singularity.\footnote{Of course, one expects that the singularity will be resolved by quantum effects. As far as the classical analyses is concerned, this is a well-defined set-up.} In \cite{Kasner:1921zz}, this has been explicitly explored and a general Kasner geometry is found to describe the singularity.

Clearly, this framework allows for a more general class of classical solutions. In particular, it allows us to pose questions about the classical interior of the black hole and its corresponding significance from the perspective of the boundary CFT. In the subsequent discussion we will address this issue, first with a toy example and then with a model that is obtained from a consistently truncated supergravity theory. For the latter, the Holographic duality is exactly known in detail.

\subsection{Construction of Flows}

Let us begin with constructing explicit flow geometries, with the action in (\ref{action1d}). At the extrema, values of the potential set a negative cosmological constant and the corresponding AdS-curvatures. This yields: $l^2_{\rm UV}=1$ and $l^2_{\rm IR}=\frac{81}{85}$. Moreover, near these extrema the second derivative of the potential sets a mass scale at the extrema i.e. $m^2=\frac{\partial^{2} V}{\partial \phi^2}\big\rvert_{\rm extremum}$. Which implies $m^2_{\rm UV}=-2$ and $m^2_{\rm IR}=2$. Correspondingly, around each extrema, the scalar deformation corresponds to turning on operators of dimensions:
\begin{equation*}
    \Delta_{\pm}=\frac{d}{2}\pm\sqrt{\frac{d^2}{4}+m^2l^2} \ ,
\end{equation*}
which gives $\Delta_{\pm}=2,1$ at UV and $\Delta_{\pm}=\frac{3}{2}\pm\frac{3}{2}\sqrt{\frac{157}{85}}$ at IR. We will turn on these operators at the UV-CFT, for a given thermal state, and explore what happens to the state under an RG-flow.

We begin with the following black hole ansatz:
\begin{equation}\label{metric1d}
    ds^2=\frac{l^2}{r^2} \left(-f(r) e^{-\chi(r)}dt^2+\frac{dr^2}{f(r)}+dx^2+dy^2\right) \ , \quad \phi=\phi(r) \ , 
\end{equation}
where $l$ is the AdS-radius. In our convention, $r \rightarrow 0$ corresponds to AdS (conformal) boundary and $r \rightarrow \infty$ corresponds to the singularity (inside the black hole). In particular, when $\phi(r) =0$, the geometry in (\ref{metric1d}) corresponds to an AdS-Schwarzschild background, with a standard singularity at $r\to \infty$. The location of the horizon is denoted by $r_{\rm H}$, with $f(r_{\rm H})=0$.  The corresponding temperature of the black hole is given by
\begin{equation}
    T=\frac{|f'_{\rm H}|e^{-\chi_{\rm H}/2}}{4\pi} \ . 
\end{equation}
Here $f'_{\rm H}=f'(r_{\rm H})$, $\chi_{\rm H}=\chi(r_{\rm H})$ and prime denotes differentiation with respect to $r$ coordinate. Furthermore, one can recast the metric in (\ref{metric1d}) in the ingoing Eddington-Finkelstein coordinates, which is regular at $r=r_{\rm H}$.

For concreteness, let us consider $d=3$. The Einstein-scalar equations are given by
\begin{eqnarray}\label{EOM1d}
r^2 f \phi''+ \left (r^2 f'-2 r f-\frac{r^2f\chi'}{2} \right )\phi'-l^2\frac{dV}{d\phi} &=0 \ , \\
r f \chi'+6f-2 r f'+2l^2 V &=0 \ , \\
\chi'-r\phi'^2 &=0    \ . 
\end{eqnarray}
The AdS-Schwarzschild solutions at the extrema are given by
\begin{eqnarray}
 && \phi(r) =0 \ , \hspace{1cm}\chi(r)=0 \ , \hspace{1cm}f(r)=l^2\left(1-\Big(\frac{r}{r_{\rm H}}\Big)^3\right) \ \label{pure AdS 1d uv} ,  \\
  && \phi(r) =\frac{2}{3} \ , \hspace{1cm} \chi(r)=0 \ , \hspace{1cm}  f(r)=\frac{85}{81} l^2 \left(1-\Big(\frac{r}{r_{\rm H}}\Big)^3 \right)   \ \label{pure AdS 1d ir} . 
\end{eqnarray}
Normalization of $f(r)$ in the above solutions \eqref{pure AdS 1d uv} and \eqref{pure AdS 1d ir} ensures that $l=l_{\rm UV}$ for the first solution and $l=l_{\rm IR}$ for the second one. \\
For a generic field configuration $\phi=\phi(r)$ the near boundary ($r\rightarrow0$) behavior of the above equations are summarized below. 
\begin{enumerate}
    \item Solution near $\phi=0$
    \begin{eqnarray}\label{sol max 1d}
   && \phi(r) =A_1 r+A_2 r^2+\cdots \ ,  \\
   &&    f(r) =l^2+\frac{A_1^2l^2}{2}r^2-f_3 r^3+\cdots \ , \\
   &&    \chi(r) =\frac{A_1^2}{2}r^2+\frac{4}{3}A_1A_2 r^ 3+\cdots \ .
        \end{eqnarray}
Normalization of $f(r)$ implies $l^2=1$ and $A_{1,2}$ are undetermined constants. The data $\{A_1, A_2\}$ set boundary conditions for the scalar field and $f_3$ determines the position of the event horizon. 

 \item Solution near $\phi=\frac{2}{3}$:
        \begin{eqnarray}\label{sol min 1d}
 &&               \phi(r)=\frac{2}{3}+B_1 r^{\frac{3}{2}+\frac{3}{2}\sqrt{\frac{157}{85}}}+B_2 r^{\frac{3}{2}-\frac{3}{2}\sqrt{\frac{157}{85}}}+\cdots \ ,  \\
  &&              f(r)=\frac{85}{81}l^2+\frac{l^2}{108}(85-\sqrt{13345})B_2^2 r^{3-3\sqrt{\frac{157}{85}}}-F_3 r^3+\cdots \ ,  \\
  &&              \chi(r) =-\frac{3}{340}B_2^2(\sqrt{13345}-85)r^{3-3\sqrt{\frac{157}{85}}}-\frac{3}{340}144 B_1B_2 r^3+\cdots \ . 
        \end{eqnarray}
Normalization of $f(r)$ implies $l^2=\frac{81}{85}$. As before, $\{B_1, B_2\}$ constitute the boundary data for the scalar field. Also, $F_3$ determines the position of the event horizon. 
\end{enumerate}
One can take either the UV-AdS or the IR-AdS, and consider turning on the scalar field around each of the extrema. The corresponding boundary data $A_{1,2}$ or $B_{1,2}$, along with the fall-off conditions of the scalar field determine the details of the CFT-deformation. It is clear from the expansions above that deformations around the UV-CFT are sourced by operators with a dimension $\Delta= 2 < d=3$. Thus, these correspond to relevant deformations. On the other hand, around the IR-CFT, the dual operators have dimension $\Delta = 3.5 > d=3$ and is thus an irrelevant deformation.

The near-singularity behaviour (inside the black hole geometry) is given by
\begin{eqnarray}
       && \phi(r) =a \log r+\cdots \ , \\
       &&  \chi(r) =\chi_0+a^2 \log r+\cdots \ , \\
       &&  f(r) =-f_0 r^{3+\frac{a^2}{2}}+\cdots \ .
\end{eqnarray}
The corresponding metric takes the form\cite{Kasner:1921zz}:
\begin{equation}\label{metricK 1d}
    ds^2=-d\tau^2+\tau^{2p_t} dt^2+\tau^{2p_x}(dx^2+dy^2) \ , \quad \phi=-\sqrt{2}p_{\phi} \log \tau \ . 
\end{equation}
Equation of motions \eqref{EOM1d} set relation between the Kasner exponents and are given by (this is done below explicitly for supergravity model):
\begin{equation}\label{Kasner rel 1d}
        2 p_x+p_t=1 \ , \quad p_t^2+2 p_x^2+2 p_{\phi}^2=1 \ .
\end{equation}
Comparing \eqref{metric1d} and \eqref{metricK 1d} and using \eqref{Kasner rel 1d}  it is easy to show the following relations:
\begin{equation}
    p_x=p_y=\frac{4}{6+a^2}\ , \hspace{1cm}p_t=\frac{a^2-2}{a^2+6} \ ,\hspace{1cm} p_{\phi}=\frac{2 a}{6+a^2} \ . 
\end{equation}

\begin{figure}[h]
	\begin{subfigure}{0.5\textwidth}
		\centering
		\includegraphics[width=\textwidth]{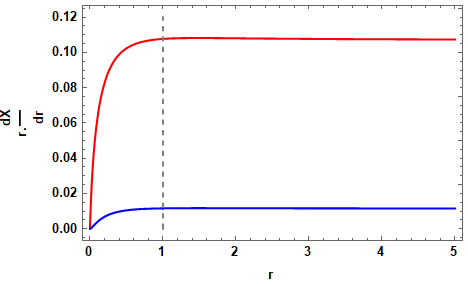}
	\end{subfigure}
	\hfill
	\begin{subfigure}{0.5\textwidth}
		\centering
		\includegraphics[width=\textwidth]{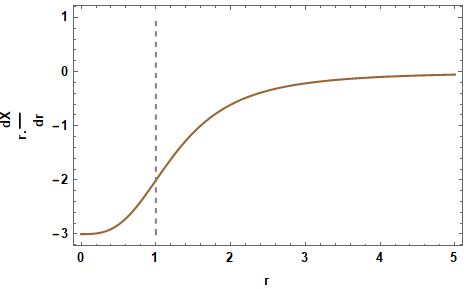}
	\end{subfigure}
	\caption{A flow from UV AdS to Kasner universe. This corresponds to a deformation parameter $\frac{A_1}{T}=6.47$. For the left figure from top to bottom $X$ are respectively $\phi(r)$ and $\chi(r)$. For the right figure $X=\log {g_{tt}'(r)}$. In the both figures dotted vertical line represents the position of horizon, $r_H$.}
	\label{fig:flow_1 one dim}
\end{figure}

\begin{figure}[h]
	\begin{subfigure}{0.5\textwidth}
		\centering
		\includegraphics[width=\textwidth]{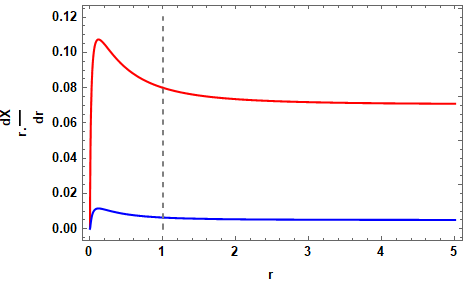}
	\end{subfigure}
	\hfill
	\begin{subfigure}{0.5\textwidth}
		\centering
		\includegraphics[width=\textwidth]{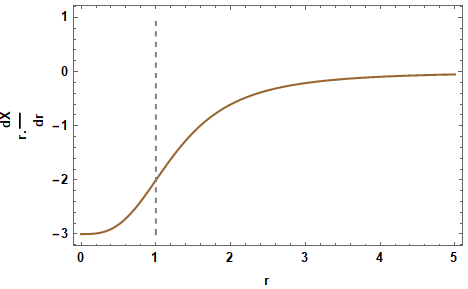}
	\end{subfigure}
	\caption{Another flow from UV AdS to Kasner universe. This corresponds to a deformation parameter $\frac{A_1}{T}=49.88$. For the left figure from top to bottom $X$ are respectively $\phi(r)$ and $\chi(r)$. For the right figure $X=\log {g_{tt}'(r)}$. In the both figures dotted vertical line represents the position of horizon, $r_H$.}
	\label{fig:flow_2 one dim}
\end{figure}

To obtain the full numerical solution of Einstein-scalar equations, we employ numerical methods. In particular, we use numerical shooting method and shoot from the horizon by imposing regularity at the horizon. The solutions are completely characterized by the value of the field at horizon and temperature, precisely by the dimensionless ratio $\frac{A_1}{T}$ (`deformation parameter'). Physically, given the UV-CFT, depending on the deformation strength set by $\frac{A_1}{T}$, we obtain a particular RG-flow of the corresponding thermal state. Generally in $(d+1)$ dimensions each of this flow solutions is characterized by a dimensionless quantity $\frac{\phi_0}{T^{d-\Delta}}$ where $\phi_0$ is boundary source term and $T$ is the temperature. Two such explicit flows are shown in Figure \ref{fig:flow_1 one dim} and Figure \ref{fig:flow_2 one dim}.

In general, for a finite deformation of the thermal state, the IR fixed point is always inside the horizon. This implies that in an equivalent Euclidean-picture, the IR fixed point is not accessible. However, in the limit $\frac{A_1}{T} \rightarrow \infty$, the IR-point approaches the horizon from inside. This happens when the scalar field takes its values at the IR-extrema close to the location of the event horizon. At the level of the action, this simply means that the scalar potential yields a corresponding negative cosmological constant.

However, AdS/CFT can be viewed as a strict statement about Holographic duality at each constant radial slice of the geometry. This is particularly sharp in the Wilsonian framework where high energy modes (correspondingly the geometric region above a constant radial slice) are integrated out and an effective action is obtained. See {\it e.g.}~\cite{Faulkner:2010jy} for an explicit construction of the same. At the UV, corresponding to $\phi=0$, we begin with a CFT. Introducing the relevant deformation triggers an RG-flow and subsequently a similar effective action can be obtained, following the treatment in \cite{Faulkner:2010jy}.\footnote{Note that, an explicit such construction may have some technical challenges, {\it e.g.}~the details in \cite{Faulkner:2010jy} depend on a scalar field with only quadratic interaction. For the toy model we have used here, the analyses need to be repeated since it involves a cubic interaction. 
The crucial point, however, is that such a construction is possible.} For any radial slice deeper into the bulk geometry, the corresponding effective action is defined. However, this procedure must stop at the event horizon, since constant radial slices outside the horizon will become constant time slices inside and one is left with only massless modes at this point. Furthermore, as we will explicitly see later, the corresponding flow equations indicate strongly that the Wilsonian effective action description remains valid only outside the event horizon, albeit infinitesimally close. See \cite{Faulkner:2010jy} for more discussions on this. It is apparent that the trans-IR flow have a different physical interpretation from the boundary CFT's perspective.

In the limit $A_1/T\to \infty$, we numerically observe that the $\phi=2/3$ extremum approaches the event horizon from inside.\footnote{Numerically it is observed that setting a value of $\phi\approx 2/3$ near the event horizon, but from inside, one is able to construct the flow solution all the way to a Kasner-singularity.} At this point, there are two possibilities: (i) The limit eventually places the $\phi=2/3$ point on the horizon or (ii) the $\phi=2/3$ point always remains inside the horizon, even though it can come arbitrarily close to it. We will momentarily elaborate on this.

\begin{figure}[h!]
    \centering
    \includegraphics[width=.8\textwidth]{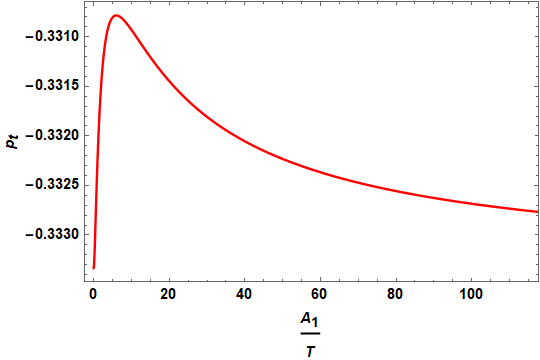}
    \caption{Plot showing how Kasner exponent $p_t$ changes with deformation $\frac{A_1}{T}$}
    \label{fig:field vs pt one dim}
\end{figure}

For any finite value of $\frac{A_1}{T}$, the IR extremum of the potential is located inside the event horizon, but it stays infinitely far from the singularity;\footnote{This is clear since the Kasner singularity behave as attractors of these equations, and therefore all details of the potential are irrelevant near them.} and the location depends on the strength of $\frac{A_1}{T}$. For each such value, there is a flow that goes all the way to the singularity which has a Kasner-structure. These geometries are characterized by the Kasner-exponents, and we have presented an exact dependence of the exponent with deformation in Figure \ref{fig:field vs pt one dim}.
It is worth emphasizing again that the IR-extremum of the potential is always separated from the singularity. This is a qualitatively similar plot to what is obtained in \cite{Frenkel:2020ysx}. The crucial difference is our potential has two extrema, and the putative IR-extremum location can be fine-tuned to approach the event horizon. We will now consider this limit more carefully.

\subsection{A Special Limiting Case}

Let us now address the question mentioned above. Physically, from the Wilsonian perspective of \cite{Faulkner:2010jy} as well as a Euclidean description of the thermal state, we expect that the exact $T=0$ flow corresponds to hitting the IR fixed point, before the flow can reach the event horizon. This way, there is no event horizon in the geometry, since the flow ends at the IR fixed point.\footnote{Note that the IR fixed point does not have any further relevant deformation.} This corresponds to the well-known flow geometries that interpolate between two AdS-vacua, {\it e.g.}~\cite{Freedman:1999gp}.

However, when we deform the thermal state with a relevant deformation, and tune $A_1/T \to \infty$, we do not expect to place the IR-extremum exactly on the event horizon. This is simply because such a configuration corresponds to the $T=0$ physics which we cannot access within the classical gravitational description, as argued on general properties of the TFD state. In other words, placing the IR-extremum on the event horizon will correspond to unentangling the TFD-state, which is not allowed. The only conclusion is therefore that the IR-extremum approaches the event horizon from inside, but never touches it. Let alone crossing it.

At the level of the equations of motion this can be further verified by constructing a series solution starting from slightly inside the event horizon. In particular, the scalar can be set to its extremum value and the corresponding solution can be continued perturbatively to higher orders. On the other hand, if we force the extremum value on the event horizon, then equations motion yield the exact solution:
\begin{eqnarray}
&& \phi= 0 \ ,  \frac{2}{3} \ , \\
&& \chi = {\rm constant} \ , \quad  f(r) = 1 - \left(\frac{r}{r_H} \right)^3 \ .
\end{eqnarray}
Clearly, the above yields two distinct AdS-Schwarzschild geometries with two different curvature. Thus, freezing the scalar to its extremum at the event horizon freezes it completely. Note that we can certainly construct a flow which begins at the UV at $\phi=0$, flows all the way and finally freezes at $\phi=2/3$ where one allows for a function $f(r)$ as above. This corresponds to a thermal state, where the black hole interior has no scalar backreaction. 
\begin{figure}[h!]
    \centering
    \includegraphics[width=.6\textwidth]{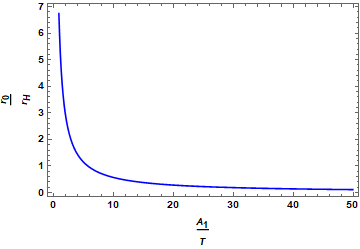}
    \caption{A pictorial representation that within the regime $r_0\sim r_H$, $A_1/T \gg 1$ can be achieved for the class of numerical solutions.}
    \label{rorhdef}
\end{figure}

Before we move on, a few remarks are in order. First, note that the gluing solution discussed above exists in the low temperature limit: $A_1/T \to \infty$. This is because the radial scale at which scalar back-reaction becomes important --- let us denote it by $r_0$ --- is far above the horizon $r_H$. This is easily understood from {\it e.g.}~the expansion in (\ref{sol min 1d}). Equating the sub-leading correction term to the leading one, we obtain: $r_0 \sim A_1^{-1}$. On the other hand, from the same asymptotic expansion, the horizon scale is given by $r_H \sim f_3^{-1/3}$. In the convention we are working, the glued solution can be constructed when $r_H \gg r_0$, which translates into the limit $A_1/T\to \infty$.

Thus, in the limit $A_1/T\to \infty$, there are two classes of solutions: One in which the scales $r_0 \sim r_H$ and the other discussed above with $r_0 \ll r_H$. Note, however, we do not expect a phase transition between these two classes of solutions, since both preserve the same symmetries as seen from the boundary CFT-dual. There are only two possibilities: Either the Euclidean free energy of the solutions with $r_0 \sim r_H$ merges into the ones with $r_0 \ll r_H$, or that the former class of solutions always remain the dominant saddle.

\section{Construction of flows in Supergravity model}

Motivated by the toy example above, let us consider a specific example that can be embedded in type IIB supergravity and therefore has a well-defined UV-completion. The dual CFT and its deformation are also explicitly known in this case. One begins with the ${\cal N}=4$ SYM which has three complex scalars, denoted by $\{\Phi_1, \Phi_2, \Phi_3\}$, in adjoint representation of the gauge group. These scalars are massless at the ${\cal N}=4$ fixed point. One can now introduce a mass-deformation for one of these scalars.\footnote{Evidently, more general mass deformations can also be considered. However, we will not discuss the most general case here.} In the ${\cal N} =1$ language, this mass deformation corresponds to deforming the superpotential by:
\begin{eqnarray}
W = {\rm Tr} \Phi_3 \left[ \Phi_1, \Phi_2\right] + \frac{m}{2} {\rm Tr} \Phi_3^2 \ ,
\end{eqnarray}
where $m$ denotes the corresponding mass. This relevant deformation induces an RG-flow and by integrating out the massive $\Phi_3$ degree of freedom, one arrives at a non-trivial interacting IR fixed point described by an ${\cal N}=1$ SYM theory\cite{Leigh:1995ep}.

Note that the vacuum state of the ${\cal N}=4$ SYM is characterized by a manifold of six real-dimensions of the corresponding marginal deformations. Specifying a set of the marginal couplings defines one particular quantization in which all states are defined. For our purpose, the structure of the TFD-state and its subsequent RG-flow is not affected by a finer description and therefore we do not include them explicitly.\footnote{We thank Nikolay Bobev for bringing this point to our attention.}

We will now construct explicit flow solutions corresponding to this flow. In particular, we will work with a five dimensional $\mathcal{N}=8$ gauged Supergravity in the bulk, whose dual is the ${\cal N}=4$ SYM theory. The corresponding potential has a richer structure compared to the toy model, but the qualitative physics is similar. The corresponding potential is given by\cite{Freedman:1999gp}:
\begin{equation}\label{pot1}
\begin{split}
 \mathcal{V} & = -\frac{g^2}{4} \Big[\rho^{-4} (1-\cos(2\phi)\big(\sinh^2(\phi_1)-(\sinh^2(\phi_2))^2\big)+\rho^2\big(\cosh(2\phi_1)+\cosh(2\phi_2)\big)\\
 &+\frac{1}{16}\rho^8 \Big(2+2\sin^2(2\phi)-2\sin^2(2\phi)\cosh(2(\phi_1-\phi_2))
 -\cosh(4\phi_1)-\cosh(4\phi_2)\Big)\Big]    \ , 
\end{split}
 \end{equation}
where $ \rho=\exp({\frac{\phi_3}{\sqrt{6}}})$ and $\phi,\phi_1,\phi_2,\phi_3$ are the scalar fields of the bulk theory. The action is given by
\begin{equation}\label{action}
    \mathcal{S}=\int d^5x \sqrt{\mid{g}\mid}\left(\frac{1}{4}R-\frac{1}{24}g^{\alpha\beta}P_{\alpha abcd}P_{\beta}^{abcd}-\mathcal{V}\right) \ .
\end{equation}
We have chosen a unit such that $\kappa_5^2=2$. The Scalar kinetic term of \eqref{action} is given by
\begin{equation}
 \frac{1}{24} g^{\mu\nu}P_{\mu abcd}P_{\nu}^{abcd} = \frac{1}{2}\left[\sum_{j=1}^3 g^{\mu\nu}(\partial_{\mu}\phi_j)(\partial_{\nu}\phi_j)\right] +\sinh^2(\phi_1-\phi_2) g^{\mu\nu}(\partial_\mu\phi)(\partial_\nu \phi) \ . 
\end{equation}
The potential \eqref{pot1} has extrema at:
\begin{equation}\label{extremum}
 (\phi,\phi_1,\phi_2,\phi_3)=(0,0,0,0) ; \left(0,\pm\frac{1}{2}\log3,0,\frac{1}{\sqrt{6}}\log2 \right) \ .     
\end{equation}

To find the nature of the extrema, we calculate the Hessian matrix constructed from the second derivatives of the potential \eqref{pot1}, which is negative semi-definite at the first point and have both positive and negative eigen values for the last two points. This implies that the first point is a local maxima and the last two points are saddle points of the potential \eqref{pot1}. As the values of $\phi$ and $\phi_2$ are always zero at the extrema, we can set this two fields to zero for our calculations (consistent with E.O.M.s). With this setting, potential\eqref{pot1} becomes:
\begin{equation}\label{pot2}
    V=-\frac{g^2}{4}\left(\rho^{-4}(1-\sinh^4\phi_1)+\rho^2(1+\cosh(2\phi_1))+\frac{\rho^8}{16}(1-\cosh(4\phi_1))\right) \ ,
\end{equation}
\begin{figure}[h!]
    \centering
    \includegraphics[width=.8\textwidth]{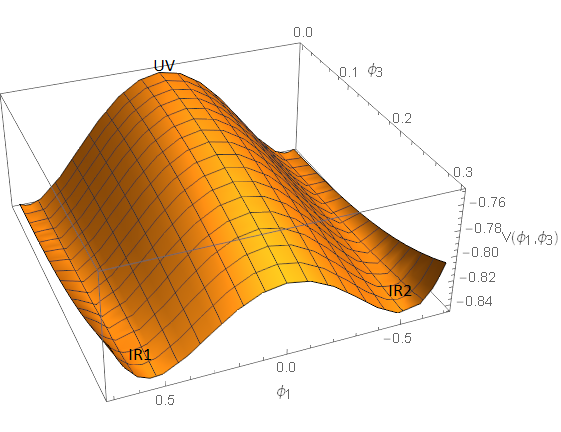}
    \caption{Image of the potential,$V$ with $\phi=\phi_2=0$.}
\end{figure}
with $(\phi_1,\phi_3)=(0,0)$ as the local maxima with $V_{\rm max}=-\frac{3g^2}{4}$ and $(\phi_1,\phi_3)=(\pm\frac{1}{2}\log3,\frac{1}{\sqrt{6}}\log2)$ as the saddle points with $V_{\rm sad}=-\frac{2^{\frac{4}{3}}g^2}{3}$. Comparing \eqref{action} with $V_{\rm max}$ and $V_{\rm sad}$ at the extrema, values of $l$ are given by $l_1=l_{\rm max}=\frac{2}{g}$ and $l_2=l_{\rm sad}=3/(2^{\frac{2}{3}}g)$. At all the extrema, values of the potential set a negative cosmological constant so an AdS solution exists at all extrema.

Usually, the maxima and the minima of the potential correspond to a UV and an IR CFT, respectively. For the potential at hand, the IR is a saddle point and contains an irrelevant direction, while the UV is a maximum and contains only relevant directions. To construct the explicit flows, let us work with the following metric ansatz:\footnote{Note that the metric ansatz can also be written in the ingoing Eddington-Finkelstein coordinates.}
\begin{equation}\label{metric1}
    ds^2=\frac{l^2}{r^2} \left(-f(r) e^{-\chi(r)}dt^2+\frac{dr^2}{f(r)}+dx^2+dy^2+dz^2\right) \ ,
\end{equation}
with $l$ denoting the AdS radius. We also assume $\phi_1=\phi_1(r)$ and $\phi_2=\phi_2(r)$. In our conventions, $r \rightarrow 0$ corresponds to AdS boundary and $r \rightarrow \infty$ is the singularity, as mentioned earlier. The horizon $r_H$ is given by the same condition $f(r=r_H)=0$. With this form of metric \eqref{metric1}, the temperature of the black hole is
\begin{equation}
    T=\frac{|f'_H|e^{-\chi_H/2}}{4\pi} \ ,
\end{equation}
which is the temperature of the dual CFT also.The Einstein-scalar equations are given by
\begin{equation}\label{EOM}
    \begin{split}
     r^2 f \phi_1''+\phi_1'\Big(r^2 f'-3 r f -\frac{r^2 f \chi'}{2}\Big)-l^2\frac{\partial V}{\partial \phi_1} &=0 \ , \\
r^2 f \phi_3''+\phi_3'\Big(r^2 f'-3 r f -\frac{r^2 f \chi'}{2}\Big)-l^2\frac{\partial V}{\partial \phi_3} &=0 \ , \\
\frac{3}{2}\Big(r f'-4f\Big)-\frac{3}{4}\chi' r f-2 l^2 V &=0 \ , \\
3 \chi'-4r\Big(\phi_1'^2+\phi_3'^2\Big) &=0   \ , 
    \end{split}
\end{equation}
where V is given by \eqref{pot2}. The AdS Schwarzschild solution at the extrema are given by
\begin{equation}\label{pure AdS}
 \begin{split}
    \phi_1 &=0 \ ,\phi_3=0\ ,\chi=0 \ ,f=\frac{(g l)^2}{4}\left(1-\Big(\frac{r}{r_H}\Big)^4\right) \ ; \\
  \phi_1 &=\pm \frac{1}{2}\log3 \ ,  \phi_3=\frac{1}{\sqrt{6}}\log2 \ ,   \chi=0 \ ,  f=\frac{2^{\frac{4}{3}}}{9} (g l)^2 \left(1-\Big(\frac{r}{r_H}\Big)^4 \right)  \ .
\end{split}   
\end{equation}
Normalization of $f$ for the above solution \eqref{pure AdS} demands that $l=l_1$ for the first solution and $l=l_2$ for the second one. For a generic field configuration $\phi_i=\phi_i(r)$ (where $i=1,3$) the near boundary ($r\rightarrow0$) behavior of the above equations are given by\\
(I) around $\phi_1=\phi_3=0$:
\begin{equation}\label{boundary expansion}
    \begin{split}
\phi_1(r) &= A_1 r+A_2 r^3 +... \ , \\
\phi_3(r) &= B_1 r^2 +2 B_2 r^2 \log r+r^4\left(B_3+B_4\log r+B_5(\log r)^2\right)... \ , \\
f(r) &= \frac{g^2 l_1^2}{4}+F_1 r^2+r^4\left(-F_2 +F_3 \log r+F_4 (\log r)^2\right)+ ... \ , \\
\chi(r) &=X_1 r^2+ r^4\left( X_2+X_3 \log r+X_4 (\log r)^2\right)+... \ . 
\end{split}
\end{equation}
Here
 \begin{equation*}
    \begin{split}
        B_3&=\frac{1}{6}\left(\sqrt{6}B_1^2-4\sqrt{6}B_1B_2+6\sqrt{6}B_2^2\right) \ ,B_4 =\frac{4}{\sqrt{6}}\left(B_1 B_2-2B_2^2\right) \ , B_5 =\frac{4}{\sqrt{6}}B_2^2 \ ; \\
        F_1 &=\frac{g^2}{6}A_1^2 \ ,F_3 =g^2\left(\frac{1}{2}B_1 B_2+\frac{2}{3} B_2^2\right) \ ,F_4 =\frac{4}{3}g^2 B_2^2 \ ; \\
        X_1 &= \frac{2 A_1^2}{3} \ , X_2 = 2A_1A_2+\frac{2}{3}\left(2B_1^2+2B_1 B_2+B_2^2\right) \ , X_3 =\frac{8}{3}\left(2 B_1B_2+B_2^2\right) \ , X_4 =\frac{16}{3}B_2^2 \ ;
    \end{split}
\end{equation*}
where $A_1,A_2,B_1,B_2,$ and $F_2$ are constants.

A comment is in order, regarding the expansion in (\ref{boundary expansion}) and the one described in \cite{Freedman:1999gp}. In (\ref{boundary expansion}) two scalar fields $\phi_{1,3}$ have independent asymptotic behaviour near the AdS-boundary. This, however, is not true in \cite{Freedman:1999gp} since the latter only considers supersymmetry preserving flows. The set of asymptotic boundary conditions in \cite{Freedman:1999gp} is therefore a subset of the ones in (\ref{boundary expansion}). Physically, we have two independent couplings that can be varied.

(II) around $\phi_1=\pm\frac{1}{2} \log 3,\phi_3=\frac{1}{\sqrt{6}}\log 2$:
\begin{equation}
\begin{split}
\phi_1(r) &=\pm\frac{1}{2}\log 3+ D_1 r^p+D_2 r^q +... \ , \\
\phi_3(r) &=\frac{1}{\sqrt{6}}\log 2\mp\frac{1+\sqrt{7}}{\sqrt{6}}\Big(  D_1 r^p+D_2 r^q\Big)+... \ ,  \\
f(r) &= \frac{2^{\frac{4}{3}}}{9}g^2 l_2^2+D_3 r^s+...  \ , \\
\chi(r) &= D_4 r^s+... \ .
\end{split}
\end{equation}
Here $p=(3-\sqrt{7}),q=4-p=(1+\sqrt{7}),s=(6-2\sqrt{7})$ and  $A_i$, $B_i (i=1,2)$ and $D_j (j=1,2,3,4)$ are the constants, where
\begin{equation}
    \begin{split}
         D_3 &= \frac{2^{\frac{10}{3}}(35-13\sqrt{7})g^2}{27(\sqrt{7}-1)(\sqrt{7}-2)}D_1^2 \ , \\
    D_4 &= \frac{8(28-11\sqrt{7})}{9(4-2\sqrt{7})}D_1^2  \ . 
    \end{split}
\end{equation}

In the interior, the near-singularity ({\it i.e.}~$r\rightarrow\infty$) behaviour is given by:
\begin{equation}\label{KasnerEq}
    \phi_1(r)=a \log r+\cdots \ , \phi_3(r)=b \log r+\cdots \ , \chi(r)= c\log r+\chi_1+\cdots \ , f(r)=-f_0 r^d+\cdots \ .
\end{equation}
Substituting these behaviours back in \eqref{EOM}, we obtain:
\begin{equation}\label{EOMsingularity}
    \begin{split}
     \frac{g^2}{16} r^{\frac{4\sqrt{6}}{3}b+4a}-\frac{1}{2} a (8+c-2d)r^d f_0 &=0 \ , \\
    \frac{g^2}{8\sqrt{6}} r^{\frac{4\sqrt{6}}{3}b+4a}-\frac{1}{2}b(8+c-2d)r^d f_0 &=0 \ , \\
    \frac{g^2}{16} r^{\frac{4\sqrt{6}}{3}b+4a}-3(8+c-2d)r^d f_0 &=0 \ , \\
   c-\frac{4}{3}(a^2+b^2) &=0   \ . 
    \end{split}
\end{equation}
There are more than one ways to solve the above set of equations. First, imagine setting: $d=\frac{4\sqrt{6}}{3}b+4a$, in which both terms on the LHS of the above equations contribute. In this case all the unknown parameter like a,b,c etc. are fixed to some particular values. However, the choice: $d> \frac{4\sqrt{6}}{3}b+4a$ gives a more general solution, which only keeps the $r^d$ term in the limit $r\to \infty$. The resulting algebraic equations can be solved for $c,d$ in terms of $a$ and $b$. This yields:
\begin{equation}\label{KasnerEq1}
    \phi_1(r)=a \log r+... \ , \phi_3(r)=b \log r+... \ , \chi(r)=\frac{4}{3}(a^2+b^2) \log r+\chi_1+... \ , f(r)=-f_0 r^{4+\frac{2}{3}(a^2+b^2)}+... \ . 
\end{equation}
Thus, near-singularity, the geometry takes a Kasner-form:
\begin{equation}\label{metricK}
    ds^2=-d\tau^2+\tau^{2p_t} dt^2+\tau^{2p_x}(dx^2+dy^2+dz^2) \ . 
\end{equation}
with $\phi_1=-\sqrt{2}p_1 \log \tau$, $\phi_3=-\sqrt{2}p_3 \log \tau$. The Einstein-scalar equations \eqref{EOM} become:
\begin{equation}\label{EOMK}
    \begin{split}
    p_t-p_x+1 &=\frac{4}{3}\frac{(p_1^2+p_3^2)}{p_x} \ , \\
    3p_x+p_t-1 &= \frac{4}{3p_x} \tau^2 l^2 V \ , \\
    3p_x+p_t-1 &= \frac{\tau^2 l^2}{p_1\sqrt{2}} \frac{\partial V}{\partial \phi_1} \ , \\
    3p_x+p_t-1 &= \frac{\tau^2 l^2}{p_3\sqrt{2}} \frac{\partial V}{\partial \phi_3} \ .
    \end{split}
\end{equation}
Consistency of the last three equations of \eqref{EOMK} implies that
\begin{equation}\label{Kcondn}
    \frac{4}{3 p_x}\tau^2 V=\frac{\tau^2}{p_1\sqrt{2}}\frac{\partial V}{\partial \phi_1}=\frac{\tau^2}{p_3\sqrt{2}}\frac{\partial V}{\partial \phi_3}=A \ ,
\end{equation}
where $A$ must be a constant. One trivial choice is of course $A=0$ and with this choice the above equations \eqref{EOMK} reduce to:
\begin{equation}
    \begin{split}
  p_t-p_x+1 &=\frac{4}{3}\frac{(p_1^2+p_3^2)}{p_x} \ , \\
    3p_x+p_t-1 &= 0    \ .
\end{split}
\end{equation}
Equation \eqref{Kcondn} can be equivalently written as: 
\begin{equation}\label{vdv}
\frac{\partial V}{\partial \phi_{1,3}}=\frac{4\sqrt{2} p_{1,3}}{3p_x}V  \ .  
\end{equation}
%

At this point, it is instructive to consider general constraints involving a supergravity potential and its gradient. A particularly interesting example is the swampland criterion of any consistent theory of quantum gravity\cite{Obied:2018sgi, Ooguri:2018wrx}:
\begin{equation}\label{bound1}
	|\nabla V|\geq \frac{c}{M_p} V \ ,
\end{equation}
or
\begin{equation}\label{bound2}
	\text{min}(\nabla_i\nabla_j V)\leq-\frac{c'}{M_p^2}V
\end{equation}
for any {\it effective} classical potential V and some positive constants $c$ and $c'$ of order one. Here $|\nabla V|=\sqrt{h_{ij}\frac{\partial V}{\partial \phi_i}\frac{\partial V}{\partial \phi_j}}$ where $h_{ij}$ is  metric on the field space defined by the kinetic term of the scalar field which is ${\rm diag}(1,1)$ for our SUGRA model and $ \text{min}(\nabla_i\nabla_j V)$ is the minimum eigenvalue of the Hessian $\nabla_i\nabla_j V $ in an orthonormal frame which is the matrix $\frac{\partial^2 V}{\partial \phi_i \partial\phi_j}$ . Using \eqref{vdv} it can be shown that 
\begin{equation*}
	|\nabla V|=\frac{|V|}{|p_x|}\frac{4}{3}\sqrt{2(p_1^2+p_3^2)} \ ,
\end{equation*}
and
\begin{equation*}
	\nabla_i\nabla_j V=\left(\frac{4\sqrt{2}}{3p_x}\right)^2 V \begin{pmatrix}
		p_1^2  & p_1p_3\\
		p_1p_3 & p_3^2
	\end{pmatrix} \ .
\end{equation*}
The eigenvalues of the above matrix are $0$ and $\left(\frac{4\sqrt{2}}{3p_x}\right)^2 V \left( p_1^2+p_3^2 \right)$. Since  ${\rm sgn}(V)$ is always negative for our model $\text{min}(\nabla_i\nabla_j V) =\left(\frac{4\sqrt{2}}{3p_x}\right)^2 V \left( p_1^2+p_3^2 \right)$.
\\ 
Then above bounds imply
\begin{align} 
	& \frac{4}{3 |p_x|}\sqrt{2(p_1^2+p_3^2)}\geq  - \frac{c}{M_p}, \ \\ 
	& \frac{32}{9}\left(\frac{p_1^2+p_3^2}{p_x^2}\right) \geq -\frac{c'}{M_p^2}.
\end{align}
The above conditions constrain the Kasner exponents further and for the SUGRA-example, these constraints are trivially satisfied.

As before, to construct the full flow solution, we can numerically solve the Einstein-scalar system of equations. The procedure is essentially the same as described in the previous section, with one physical distinction. For each set of horizon data and a corresponding numerical solution, we will obtain two independent coupling at the boundary, corresponding to two real-valued deformations of the ${\cal N}=4$ SYM. This will become pictorially clear momentarily.

Note that, near each critical point, the full potential in \eqref{pot2} takes a local form: $V(\phi) = \Lambda + m^2 \phi^2$, which is identical to the potential considered in \cite{Frenkel:2020ysx}. The main difference, however, is that there is now no free parameter ({\it i.e.}~$m^2$ is not adjustable at will, rather it is completely fixed by the supergravity theory). For our example, the mass matrix around the UV-extremum corresponds to two relevant operators. Therefore, any UV-deformation will induce an RG-flow much like the ones considered in \cite{Frenkel:2020ysx}. The mass matrix around the IR-extremum, on the other hand, corresponds to one relevant and one irrelevant deformation and there is no further truncation in which only the relevant deformation can be turned on. Thus, once the IR-CFT is reached, there can no longer be a further flow.\footnote{This statement is strictly true for the supersymmetric flows with the two-scalar truncation of the full supergravity potential. Relaxing this will allow one to construct much richer possibilities which are outside the scope of the current work. We thank Nikolay Bobev for raising this point to us.} For a detailed account of the dimension of operators, see Appendix \ref{Appendix:relevant}.
\begin{figure}[h]
	\begin{subfigure}{0.5\textwidth}
		\centering
		\includegraphics[width=\textwidth]{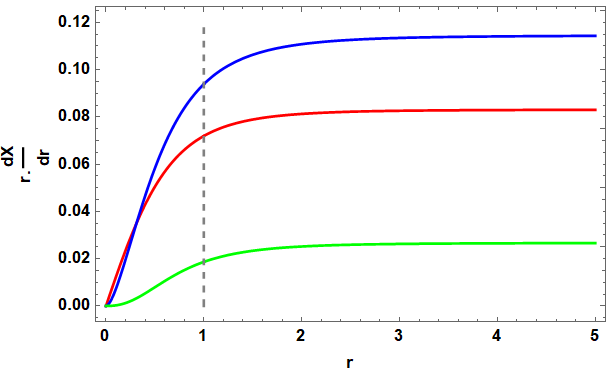}
	\end{subfigure}
	\hfill
	\begin{subfigure}{0.5\textwidth}
		\centering
		\includegraphics[width=\textwidth]{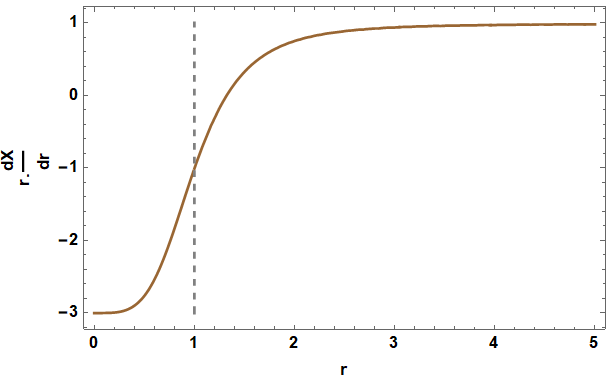}
	\end{subfigure}
	\caption{A flow from AdS to Kasner universe about the maximum of V. This corresponds to deformation parameter $\frac{A_1}{T}=0.38$. For the left figure from top to bottom $X$ are respectively $\phi_3(r)$, $\phi_1(r)$ and $\chi(r)$. For the right figure $X=\log {g_{tt}'(r)}$. In the both figures dotted vertical line represents the position of horizon, $r_H$.}
	\label{fig:flow_(0,0)}
\end{figure}
Thus, all non-trivial flows, between an AdS at $r\to 0$ boundary, to a Kasner Universe at $r\to \infty$ must not reach the IR extremum outside the event horizon.\footnote{As we just argued, if the IR extremum of the potential is reached outside the event horizon, the system cannot flow any further, since the model does not accommodate a single relevant deformation at this point.} Therefore, the IR extremum of the potential is reached only inside the event horizon, or at the limiting case, can approach event horizon. As we have discussed with the toy model, freezing the scalars to their IR-extremum values at the horizon freezes the flow completely. In the first case, the Euclidean section of the geometry never has the access to the IR fixed point and the flow ends at the horizon, even in the limit of vanishing temperature. In the latter case, which is only relevant at very small temperatures, one obtains an AdS-Rindler patch at the horizon and the corresponding Euclidean section touches the IR fixed point. However, as we have also discusses, there is no phase transition between these two classes since they both preserve the same symmetry. This can be further supported by computing the Euclidean free energies which prefers the former class of solutions.

\begin{figure}[h]
\begin{subfigure}{0.6\textwidth}
    \centering
    \includegraphics[width=\textwidth]{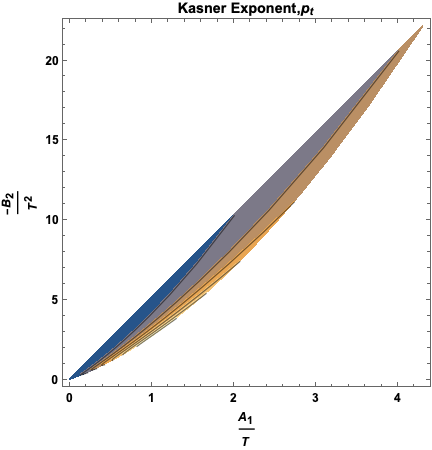}
    \end{subfigure}
    \hfill
    \begin{subfigure}{0.07\textwidth}
    \includegraphics[width=\textwidth]{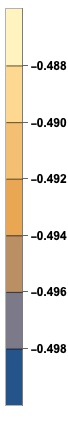}
    \end{subfigure}
    \caption{A contour plot showing how the Kasner exponent depends on the deformation parameters. This is obtained from the explicit numerical flow solutions constructed with the supergravity potential. Qualitatively, the Kasner exponent depends on the deformation in a similar manner that we have observed with the toy model.}
    \label{fig:2ptsugra}
\end{figure}

In Figure \ref{fig:flow_(log,log)} we have shown numerically obtained flows for moderately large deformation and the behavior of the scalar fields and metric functions for the above flow is given in Figure \ref{fig:fields_(log,log)}.
\begin{figure}[t]
	\begin{subfigure}{0.5\textwidth}
		\centering
		\includegraphics[width=\textwidth]{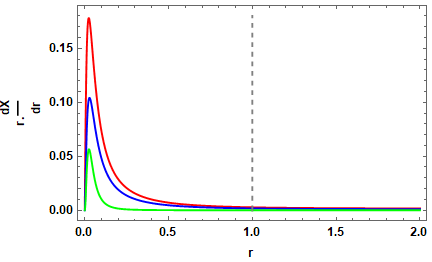}
	\end{subfigure}
	\hfill
	\begin{subfigure}{0.5\textwidth}
		\centering
		\includegraphics[width=\textwidth]{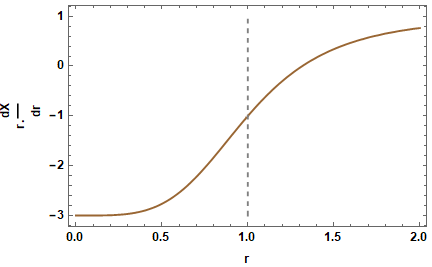}
	\end{subfigure}
	\caption{A flow from AdS to Kasner universe for $\phi_1(r_h)=\frac{1}{2}\log3-0.001,\phi_3(r_h)=\frac{1}{\sqrt{6}}\log2-0.001$. This corresponds to deformation $\frac{A_1}{T}=50.98$. For the left figure from top to bottom $X$ are respectively $\phi_1(r)$, $\phi_3(r)$ and $\chi(r)$. For the right figure $X=\log {g_{tt}'(r)}$. In the both figures dotted vertical line represents the position of horizon, $r_H$.}
	\label{fig:flow_(log,log)}
\end{figure}
\begin{figure}[h!]
	\begin{subfigure}{0.5\textwidth}
		\centering
		\includegraphics[width=\textwidth]{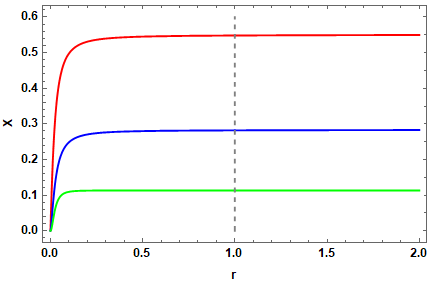}
	\end{subfigure}
	\hfill
	\begin{subfigure}{0.5\textwidth}
		\centering
		\includegraphics[width=\textwidth]{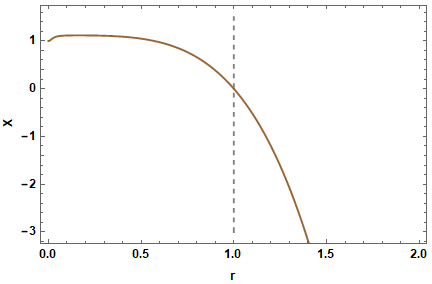}
	\end{subfigure}
	\caption{Behaviour of fields and metric functions for the flow of Figure \ref{fig:flow_(log,log)}. In the left figure from top to bottom $X$ are respectively $\phi_1(r)$, $\phi_3(r)$ and $\chi(r)$. For the right figure $X=f(r)$. In the both figures dotted vertical line represents the position of horizon, $r_H$.}
	\label{fig:fields_(log,log)}
\end{figure}
As mentioned before, there are two independent dimensionless deformations: $\frac{A_1}{T}$ for $\phi_1$ and $\frac{-B_2}{T^2}$ for $\phi_3$ for this problem. In Figure \ref{fig:2ptsugra} we have shown how the Kasner exponent $p_t$ depends on them. 
Note that, near the singularity, matter-dependence washes out and we are left with an attractor-type mechanism with a Kasner universe. The details of the supergravity potential, in particular the explicit CFT-dual and its specific deformations, do not appear important for this structure. Nonetheless, the details remain imprinted on the maximum of $p_t$, which explicitly depends on the number of dimensions and the explicit matter interaction in the gravitational description. This imprint, in turn, leaves a subtle signature on the corresponding two-sided correlation functions that we will explicitly present later.

\section{Flow of the Coupling Constant}

In \cite{Faulkner:2010jy} an explicit flow equation for the relevant deformation was also obtained. In this section, we use these flow equations to further sharpen the statement of why a standard RG-flow interpretation should end at the horizon. For this, we will simply solve the flow equations, in a black hole geometry, corresponding to single trace and double trace deformations that are discussed at length in \cite{Faulkner:2010jy} when the horizon vanishes or becomes extremal. In this section we will review some basic features of \cite{Faulkner:2010jy} and end with a couple of comments supporting the idea that the RG-flow terminates at the event horizon.

One begins by considering a bulk action with scalar field $\phi$ of the form:
 \begin{equation}
    S=\int_{r>\epsilon} d^{d+1}x \sqrt{-g}\mathcal{L}(\phi,\partial_M\phi)+S_B[\phi,\epsilon] \ , 
\end{equation}
where $\mathcal{L}$ is the Lagrangian in the bulk and the boundary action $S_B$ is defined at $r=\epsilon$ which corresponds to some UV cutoff $\Lambda$ in the field theory. The dual QFT path integral can be described by a path integral:
 \begin{equation}
    Z=\int_\Lambda D\Phi \, {\rm exp} [i I_{\rm eff}[\Phi,\Lambda]] \ , 
\end{equation}
where $\Phi$ collectively denotes all the fields in the QFT and all high energy modes above $\Lambda$ have been integrated out in obtaining the effective action $I_{\rm eff}$. The expansion of boundary action $S_B$ in momentum space can be written as:
\begin{equation}
    S_B[\phi,\epsilon]=\Lambda(\epsilon)+\int\frac{d^dk}{(2 \pi)^d}\sqrt{-\gamma}J(k,\epsilon)\phi(-k)-\frac{1}{2}\int \frac{d^dk}{(2 \pi)^d}\sqrt{-\gamma}f(k,\epsilon)\phi(k)\phi(-k) \ .
\end{equation}
Here $\gamma$ is the induced metric at $r=\epsilon$ slice. Also $J(k)$ and $f(k)$ are related to couplings of single trace and double trace operators defined in terms of a local, gauge-invariant operator $\mathcal{O}$, where $\mathcal{O}$ is the CFT dual to the bulk field $\phi$.  As discussed there, the condition that on shell action is independent of the cutoff $r=\epsilon$ gives a set of flow equations which, for the potential $V(\phi)=\frac{1}{2}m^2\phi^2$, reduce to:
    \begin{align}
        &\frac{1}{\sqrt{-g}}\partial_\epsilon \Lambda =\frac{1}{2}\int \frac{d^dk}{(2\pi)^d}J(k,\epsilon)J(-k,\epsilon) \ , \\
        &\frac{1}{\sqrt{-g}}\partial_\epsilon (\sqrt{-\gamma}J(k,\epsilon))=-J(k,\epsilon)f(k,\epsilon) \ , \\
        &\frac{1}{\sqrt{-g}}\partial_\epsilon (\sqrt{-\gamma}f(k,\epsilon))=-f^2(k,\epsilon)+k^{\mu}k_{\mu}+m^2 \ .
    \end{align}

Our goal here is to solve these flow equations in a black hole background of the form:
\begin{equation}
    ds^2=\frac{1}{r^2}\Big(-h(r)dt^2+d\Vec{x}^2 \Big)+\frac{1}{r^2}\frac{dr^2}{h(r)} \ . 
\end{equation}
Here $r$ is the holographic coordinate i.e. related to energy scale of the dual field theory. For $AdS_{d+1}$ $h(r)=1-(\frac{r}{r_H})^d$ and Hawking temperature $T_H=\frac{d}{4\pi r_H}$ which is the temperature of the dual field theory. Here we consider only the outside of the horizon which we have set at $r_H=1$; also we focus on zero momentum sector of a massless scalar field ($ k_{\mu}=0,m=0$) for simplicity. In this case, the solutions of $f(r)$ and $J(r)$ are given by,
\begin{align}
    f(\epsilon) &=\frac{d \epsilon^d}{\sqrt{1-\epsilon^d}( \mathcal{F}_0 d-\log(1-\epsilon^d))} \ , \label{sol1} \\
    J(\epsilon) &=\frac{\mathcal{J}_0 \epsilon^d}{\sqrt{1-\epsilon^d}( \mathcal{F}_0 d-\log(1-\epsilon^d))} \ .  \label{sol2}
\end{align}
Here $\mathcal{F}_0$ and $\mathcal{J}_0$ can be written in terms of initial conditions $f_0,J_0$ at $r=\epsilon_0$ as $\mathcal{F}_0=\epsilon_0^d f_0^{-1}$ and $\mathcal{J}_0= J_0 d f_0^{-1}$. The explicit temperature dependence can be restored by replacing $\epsilon \to \frac{4\pi T_H}{d}\epsilon$, following dimensional analysis. As $\epsilon \to 1$, {\it i.e.}~the cut-off surface is taken closer to the horizon, both $f$ and $J$ diverge. The deformations become infinitely large, and any dynamics is essentially frozen at the horizon. Thus, physically, the RG-flow should end here.

We end this brief discussion with a remark. While it is indeed true that the conventional RG-flow cannot go beyond the classical event horizon, there is a sense in which the ``RG-flow" can be continued across it and into the black hole interior\cite{Caceres:2022smh}. This structure is completely generic and applies to black hole interiors with and without a scalar back-reaction. It will be interesting to explore this aspect in our case as well.

\section{Correlations under an RG}

In this section, we will compute correlations between the left and the right degrees of freedom in the dual boundary CFT. We will do this by first computing two-point functions between heavy operators of the thermofield double state of the dual CFT, with two operators inserted at the two boundaries in the corresponding Penrose diagram\cite{Maldacena:2001kr}. We will also calculate entanglement between the left and the right degrees of freedom by considering two copies of half of space on two sides of the TFD state, as the corresponding subregion.

\subsection{Two-point Correlation Function}

Let us consider a correlator with operators with large conformal dimensions. The essential technicalities are discussed in many places and we will closely follow the conventions in \cite{Frenkel:2020ysx}. In the gravitational description, this is equivalent to calculating spacelike geodesic length between two asymptotic AdS regions in the Kruskal patch. For spacelike geodesic:
\begin{equation}\label{geodesic}
     g_{tt}\dot{t}^2+g_{rr}\dot{r}^2=1 \ .
\end{equation}
Here dot denotes a derivative with respect to proper time $\tau$. Time translation symmetry of the metric \eqref{metric1} implies that there exists a conserved quantity $E$ (energy), given by $ E=-g_{tt}\dot{t}$. With this equation \eqref{geodesic} becomes:
  \begin{align}\label{geodesic1}
      &\frac{E^2}{g_{tt}}+g_{rr}\dot{r}^2=1 \ , \\
      &\dot{r}=\pm\sqrt{\frac{1}{g_{tt}}\left(1-\frac{E^2}{g_{tt}}\right)} \ . 
  \end{align}
For ingoing radial geodesic we get:
\begin{equation*}
    \frac{dt}{dr}=-\frac{\sqrt{-g_{tt}g_{rr}}}{g_{tt}\sqrt{E^2-g_{tt}}}E \ .
\end{equation*}  
To include some information about the deformed singularity ({\it i.e.}~the Kasner region) in the correlation function, the geodesic has to probe well inside the horizon, which requires very large energy. In the limit $E\rightarrow\infty$ turning point (where $\dot{r}$ is zero) is well inside the horizon and given by 
\begin{equation*}
     E^2 =g_{tt}(r_\ast)=-\frac{f(r_\ast)e^{-\chi(r_\ast)}}{r_\ast^2} \ .
\end{equation*}
Using the behaviour \eqref{KasnerEq1} of metric functions we get:
\begin{equation}\label{turning point}
    r_\ast=\left(\frac{E^2}{f_0 e^{-\chi_1}}\right)^{\frac{1}{d-c-2}} \ .
\end{equation}

The boundary time for a radial spacelike geodesic of energy $E$ to reach $r_\ast$ is given by
\begin{equation}
    \begin{split}
        t(r_\ast)-t(0) &=- \int_{0}^{r_\ast} \frac{\sqrt{-g_{tt}g_{rr}} E}{g_{tt}\sqrt{E^2-g_{tt}}} \,dr \ , \\
        &=\int_{0}^{r_\ast} \frac{{\rm sgn}(E)  e^{\frac{\chi}{2}}}{f\sqrt{1+\frac{f e^{-\chi}}{(r E)^2}}} \,dr \ , \\
        &=P\int_{0}^{r_\ast} \frac{{\rm sgn}(E)  e^{\frac{\chi}{2}}}{f\sqrt{1+\frac{f e^{-\chi}}{(r E)^2}}} \,dr +\frac{i}{4T} \ . 
    \end{split}
\end{equation}
For a symmetric geodesic, the real part of $t(r_\ast)$ is zero, and so boundary time $t(0)$ for this geodesic is given by
\begin{equation}\label{boundary time}
    t(0)=-P\int_{0}^{r_\ast} \frac{{\rm sgn}(E)  e^{\frac{\chi}{2}}}{f\sqrt{1+\frac{f e^{-\chi}}{(r E)^2}}} \,dr \ .
\end{equation}
From equation \eqref{geodesic1} the regulated length for such geodesic is:
\begin{equation}\label{geodesic length}
    \begin{split}
        L &= 2\int_{r_c}^{r_\ast} \frac{\sqrt{-g_{tt}g_{rr}} }{\sqrt{E^2-g_{tt}}} \,dr +\frac{4}{g l}\log r_c \ , \\
        &=\frac{2}{|E|}\int_{r_c}^{r_\ast} \frac{e^{\frac{-\chi}{2}}}{r^2\sqrt{1+\frac{f e^{-\chi}}{(r E)^2}}} \,dr+\frac{4}{g l}\log r_c \ . 
    \end{split}
\end{equation}
The above integrations \eqref{boundary time} and \eqref{geodesic length} are somewhat subtle and their evaluations are given in Appendix \ref{Appendix:Integration}. The main results are given below:
\begin{equation}\label{boundary time final}
    t(0)=t_0+\frac{T_1}{E}+\frac{t_2}{E^2}+\frac{T_3}{E^3}+\frac{t^4}{E^4}+\frac{T_5}{E^5}+T_6\frac{\log E}{E^5}+T_7\frac{(\log E)^2}{E^5}+\frac{T_8}{E^{\frac{12}{4-c}}}+\cdots  \ . 
\end{equation}
\begin{equation}\label{geodesic length final}
    L=L_0+
    \frac{4}{g l}\log\Big(\frac{2}{E}\Big)+\frac{l_1}{E}+\frac{L_2}{E^2}+\frac{l_3}{E^3}+\frac{L_4}{E^4}+L_5\frac{\log E}{E^4}+L_6\frac{(\log E)^2}{E^4}+\frac{L_7}{E^{\frac{c+8}{4-c}}}+\cdots \ .
\end{equation}
Here the values of $t_0,t_2,t_4,l_1,l_3,\cdots$ depend on the behavior of the metric functions. Relevant detailed expressions are relegated to appendix D. To obtain the geodesic length $L$ in terms of boundary time $t(0)$, let us assume $\Delta t=|t(0)-t_0|$, then expression \eqref{geodesic length final} becomes:
\begin{equation}\label{correlation function}
    \begin{split}
        L=&L_0+\frac{4}{g l}\log(2\Delta t)+c_1 \Delta t+c_2(\Delta t)^2+c_3(\Delta t)^3+c_4(\Delta t)^4+\\
        &L_5(\Delta t)^4\log(\Delta t)+L_6(\Delta t)^4(\log(\Delta t))^2+L_7 (\Delta t)^{-\frac{1}{p_t}}+\cdots \ . 
    \end{split}
\end{equation}
From equation \eqref{correlation function}, it is clear that correlation function of this type contains information about the singularity through $p_t$ (recall that Kasner universe is characterized by $p_t$ and $p_x$). So far, these features are qualitatively similar to the ones discussed in \cite{Frenkel:2020ysx}, while details depend on the specific model in a specific number of dimensions. 
\begin{figure}[h!]
    \centering
    \includegraphics[width=.6\textwidth]{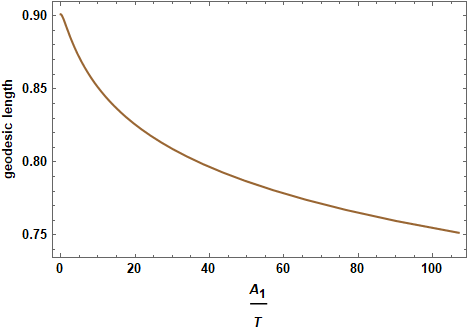}
    \caption{A representative plot of the renormalized geodesic length connecting the end points on the right and the left boundaries, where the corresponding operator is localized. In this particular case, we have set the boundary time $t_{\rm b}=0$ for convenience. We have shown the dependence of the renormalized geodesic length as a function of the deformation parameter, {\it a la} equation (\ref{pure AdS}) in the toy model.}
    \label{2pttoy}
\end{figure}
\begin{figure}[h]
\begin{subfigure}{0.6\textwidth}
    \centering
    \includegraphics[width=\textwidth]{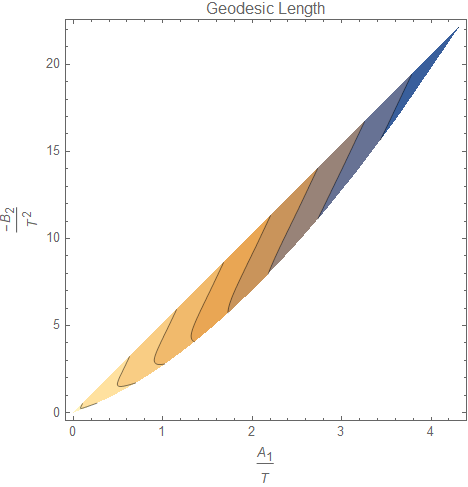}
    \end{subfigure}
    \hfill
    \begin{subfigure}{0.07\textwidth}
    \includegraphics[width=\textwidth]{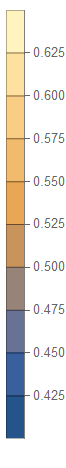}
    \end{subfigure}
    \caption{A representative contour plot of the renormalized geodesic length connecting the end points on the right and the left boundaries, where the corresponding operator is localized. In this particular case, we have set the boundary time $t_{\rm b}=0$ for convenience. We have shown the dependence of the renormalized geodesic length as a function of the deformation parameters, {\it a la} equation (\ref{boundary expansion}) in the supergravity model.}
    \label{2ptsugra}
\end{figure}

The equal-time two-point function is obtained from the renormalized geodesic length, by:
\begin{eqnarray}
\left\langle {\cal O}_{\rm L}(t, x) {\cal O}_{\rm R}(t,x)\right \rangle = e^{- \Delta {\cal L}_{\rm ren}} \ , \label{2ptgen}
\end{eqnarray}
where $\Delta$ is the dimension of the heavy operator ${\cal O}_{\rm L,R}$ and ${\cal L}_{\rm ren}$ is the renormalized geodesic length connecting the operators on two sides. Given all details, the geodesic approximation yields a functional dependence on how this two-point function depends on time. In figure \ref{2pttoy}, we have presented an explicit dependence of the geodesic length with the deformation parameter, at fixed boundary time $t_{\rm b}=0$, using the toy model. As the figure suggests, renormalized geodesic length decreases with increasing deformation and therefore by (\ref{2ptgen}), the corresponding two-sided correlator increases at a fixed time-slice on the boundary. This behaviour is similar to that of an one-sided two-point correlator, as a function of the distance between the operators.  A similar behaviour is further supported by an analogous calculation with the supergravity potential in section $4$, which is summarized in figure \ref{2ptsugra}. In this case, the renormalized geodesic length monotonically decreases along both deformations, thereby enhancing the two-sided two point correlator.

\subsection{Entanglement Entropy}

Entanglement is an extremely important probe for a state, particularly so for the TFD-state which is a maximally entangled one. Guided by the results in the previous subsection, we expect that entanglement between the two CFTs also decreases as the deformation is increased, although the state remains maximally entangled. In this subsection we will explicitly compute entanglement entropy to further establish the same. In Holography, entanglement entropy is computed using the well-known Ryu-Takayanagi prescription in \cite{Ryu:2006bv, Hubeny:2007xt}, by calculating the area of a co-dimension two extremal hypersurface, anchored at a specific boundary region at the conformal boundary of AdS. For static geometries, this calculation is performed at a constant boundary time-slice, which defines the Hilbert space of the boundary theory and its factorization in terms of a sub-region and its complement.

Let us therefore consider a constant boundary time-slice, and define a sub-region as the half-space on both sides of the TFD-state. The corresponding Ryu-Takayanagi surface is parametrized by a curve $t(r)$ in the $\{t, r\}$-submanifold of the background. This extremal area surface is similar to the spacelike geodesic described in the previous section. The only difference is that the turning point for extremal are surfaces does not approach the singularity even at late times. We will not present the details of this calculation, since it follows closely the standard Hartman-Maldacena computations in \cite{Hartman:2013qma} and for our specific purposes the relevant framework already appears in \cite{Frenkel:2020ysx, Caceres:2021fuw}. 
\begin{figure}[t]
    \centering
    \includegraphics[width=.6\textwidth]{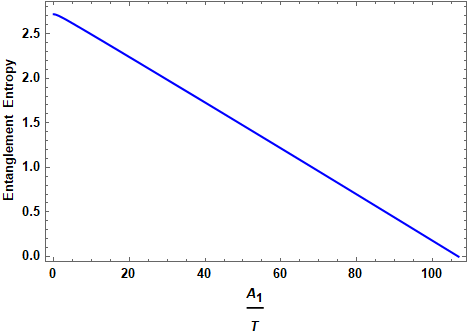}
    \caption{A representative plot of the entanglement entropy, as defined by $\Delta S$, for a sub-region that consists of two copies of the half-space on two sides of the TFD-state. In this particular case, we have set the boundary time $t_{\rm b}=0$ for convenience. We have shown the dependence of the renormalized entanglement entropy as a function of the deformation parameter, {\it a la} equation (\ref{pure AdS}) in the toy model.}
    \label{eetoy}
\end{figure}
\begin{figure}[t]

    \centering
    \includegraphics[scale=0.75]{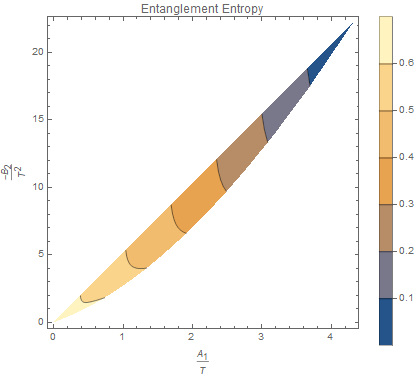}
    
     \caption{A representative contour plot of the entanglement entropy for a sub-region that consists of two copies of the half-space on two sides of the TFD-state. In this particular case, we have set the boundary time $t_{\rm b}=0$ for convenience. We have shown the dependence of the renormalized entanglement entropy  as a function of the deformation parameters, {\it a la} equation (\ref{boundary expansion}) in the supergravity model.}
    \label{eesugra}
\end{figure}
For simplicity, let us work with the boundary Cauchy slice at $t_{\rm b}=0$. In this case, the area of the Hartman-Maldacena surface, in an asymptotically AdS$_{d+1}$-background, can be computed by evaluating the following integral\cite{Caceres:2021fuw}:
\begin{eqnarray}
{\cal A}_{\rm HM} = 2 \int_0^{r_*} \frac{dr}{r^{d-1} \sqrt{f(r)}} \ ,
\end{eqnarray}
where $r_*$ is the location of the turning point, which is determined by solving $f(r_*) e^{-\chi(r_*)} = 0$, for $t_{\rm b}=0$. This area is divergent near the conformal boundary, which we can regulate by subtracting an appropriate counter-term. For our purposes, though, we find it convenient to compute the following difference: $\Delta S = S(\phi_0/T) - S(\infty)$, which automatically cancels the divergences.\footnote{Specifically at late boundary times, entanglement entropy grows linearly in time. This growth can be quantified by an entanglement velocity defined as: $v \sim \partial {\cal A}_{\rm HM}/ \partial t_{\rm b} $. We have focussed on the $t_{\rm b}=0$ slice for simplicity, and therefore we will not explore the behaviour of the entanglement velocity, although numerical investigations suggest that, defined in the usual manner, the corresponding entanglement velocity has a qualitatively similar behaviour to \cite{Frenkel:2020ysx, Caceres:2021fuw}.}

The numerical results are described in figures \ref{eetoy} and \ref{eesugra}. In both of them, we have shown the dependence of $\Delta S$, defined above, as a function of $\phi_0/T^{d-\Delta}$. Clearly, the corresponding entanglement entropy is monotonically decreasing, in keeping with the intuition of coarse-graining of the RG-flow. It is curious to note that the behaviour seems linear for a large regime that may have a simple underlying reason which we will not explore here. Moreover, in both the figures, entanglement entropy appears to vanish sharply at a sufficiently large value of $\phi_0/T^{d-\Delta}$. This is only an artefact of placing a numerical cut-off for the limit $\phi_0/T^{d-\Delta} \to \infty$ and we expect the curves to actually slowly level off.

\section{Holographic $a$ function and its behavior}

At zero temperature there is a well-established, unambiguous notion of a UV and an IR fixed point. It is clear from the earlier sections that for finite temperature there are ambiguities in the identification of the fixed points, especially in the IR. If one defines the IR fixed-point as the point where the scalars take the value of their corresponding zero-temperature IR fixed point, then this point can not be located outside the horizon. We have observed this to be true for both the toy model and the SUGRA model, in the previous sections. One may intuit the horizon as the IR fixed point, but the flows do not stop there. So, a natural question is what physics is encoded in the trans-IR flow. By construction, the trans-IR region captures the black hole interior. In \cite{Caceres:2022smh}, a corresponding holographic $a$-function was proposed for the entire flow:
\begin{equation}\label{expression of a}
    a_{\beta}=\frac{\pi^{d/2}}{ \Gamma \big(\frac{d}{2}\big) l_P^{d-1}} {\rm exp} \Big[-\frac{d-1}{2}\chi(r)\Big] \ ,
\end{equation}
where $\beta$ is the inverse temperature of the black hole as well as the dual field theory and $l_P$ denotes Planck length. Null energy condition ensures the monotonicity of this $a$-function along the flow, including the trans-IR region. It matches with the value of conventional $a$-function at the UV boundary.

So we explore where the $a$-function reaches the value of conventional IR fixed point (let us denote this point as $r_{\rm IR}$). We address this question in the following and, as expected, this position is a function of the deformation. For the toy model $r_{\rm IR}$ is always inside the horizon for any values of the deformation and for SUGRA model $r_{\rm IR}$ can be pulled outside the horizon for large enough deformations.

\subsection{Toy Model}

\begin{figure}[t]
    \centering
    \includegraphics[width=.6\textwidth]{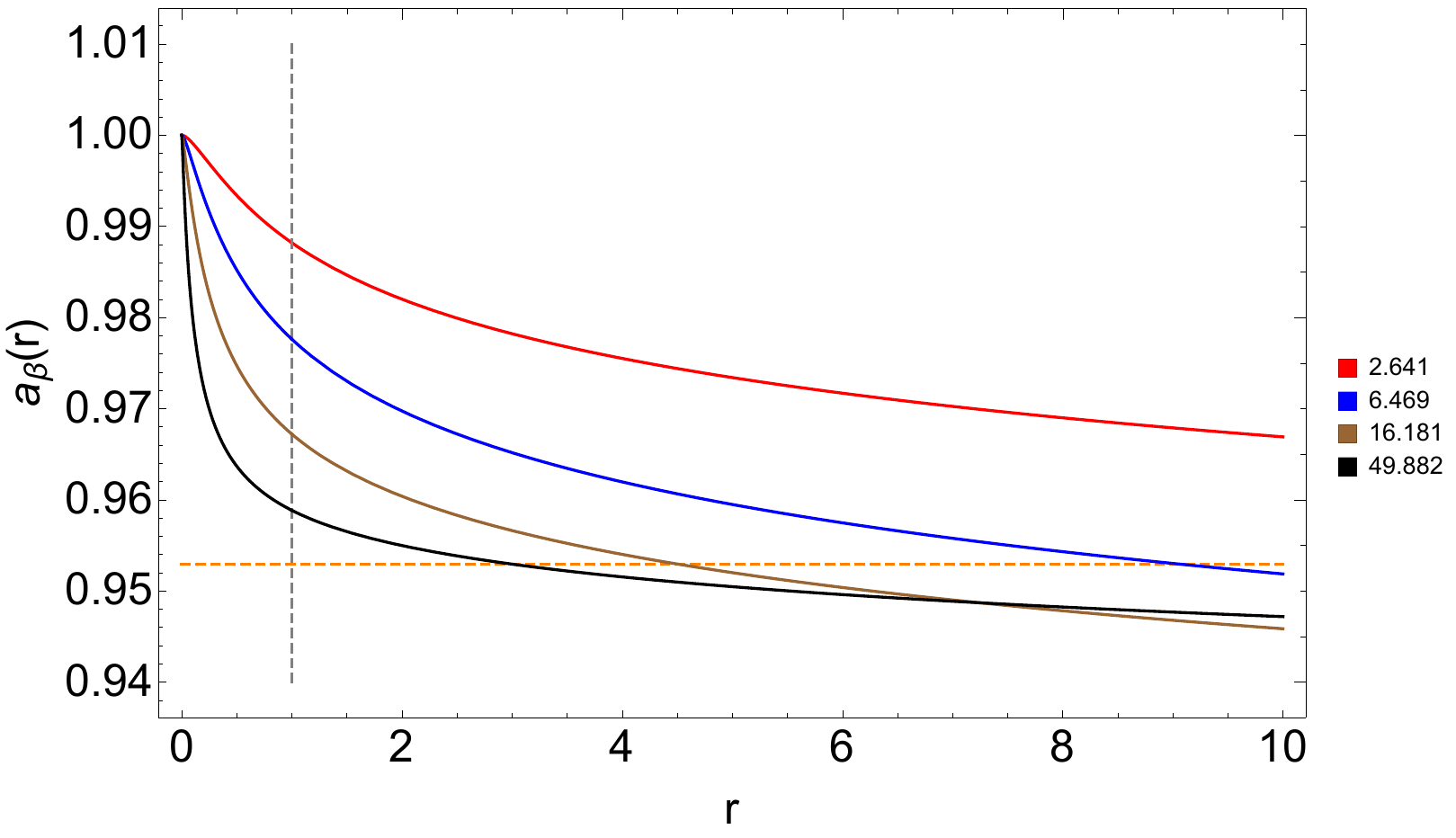}
    \caption{Plot of $a_{\beta}$ (in the unit of $2\pi l_p^{-2}$) vs $r$, for the toy model. From top to bottom, the deformation increases (values of $\frac{A_1}{T}$ are shown in the right). The vertical gray line represents the position of the horizon and the horizontal orange line represents $a_{\beta}=\frac{81}{85}$.}
    \label{cwithr}
\end{figure}

\begin{figure}[h!]
    \centering
    \includegraphics[width=.6\textwidth]{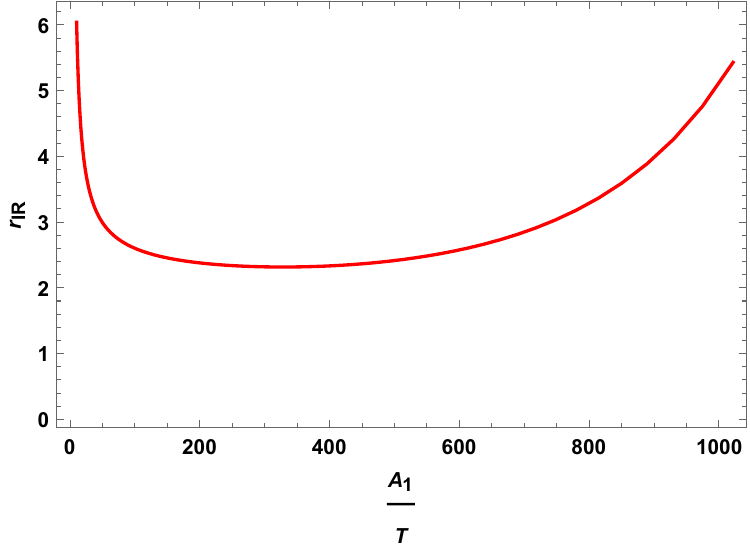}
    \caption{Plot of $r_{\rm IR}$ with the deformation parameter $\frac{A_1}{T}$,  for the toy model.}
    \label{cwithdeformation}
\end{figure}

For the toy model ($d=3$) expression for $a_{\beta}$ \eqref{expression of a} becomes:
\begin{equation*}
    a_{\beta}=\frac{\pi^{3/2}}{\Gamma\left(\frac{3}{2}\right) l_P^2} {\rm exp}[-\chi(r)] \ . 
\end{equation*}
The ratio of conventional $a$-function for the toy model is $\frac{a_{\rm IR}}{a_{\rm UV}}=\frac{81}{85}$ \footnote{For $AdS_{d+1}$, $a$-function of the boundary $CFT_d$, $a \sim l^{d-1}$, where $l$ denotes AdS radius.} and we want to keep track of $r_{\rm IR}$ where it acquires this ratio. A plot of $a_{\beta}$ with r for different deformation parameters is shown in Figure \ref{cwithr}. It is a monotonically decreasing function as is guaranteed.
Figure \ref{cwithdeformation} clearly demonstrates at which $r=r_{\rm IR}$ the above ratio is obtained, as a function of the deformation parameter $\frac{A_1}{T}$. For small $\frac{A_1}{T}$, $r_{\rm IR}$ is well inside the horizon. As $\frac{A_1}{T}$ increases it comes close to the horizon and then bounces back towards the singularity.
%

\subsection{Supergravity Model}
\begin{figure}[h]
    \centering
    \includegraphics[width=.7\textwidth]{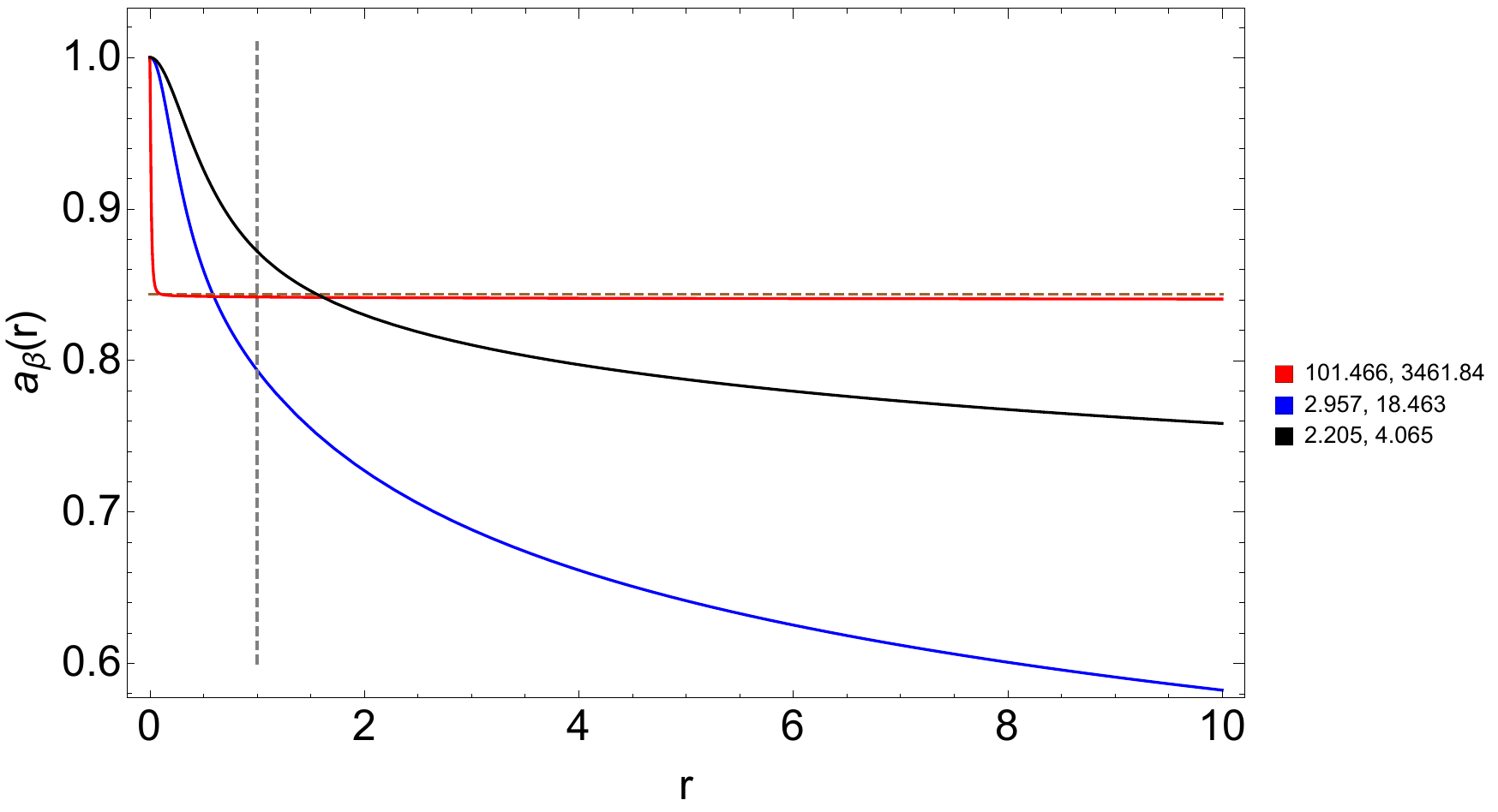}
    \caption{Plot of $a_{\beta}$ (in the unit of $2\pi^2 l_p^{-3}$) vs $r$ for the SUGRA model. We have chosen three different set of deformations (values of deformations ($\frac{A_1}{T}$, $\frac{-B_2}{T^2}$), shown on the right). The vertical line represents the position of the horizon and the horizontal line represents $a_{\beta}=\frac{27}{32}$.}
    \label{cwithdeformationSUGRA}
\end{figure}

\begin{figure}[h]
\begin{subfigure}{0.5\textwidth}
    \centering
    \includegraphics[width=\textwidth]{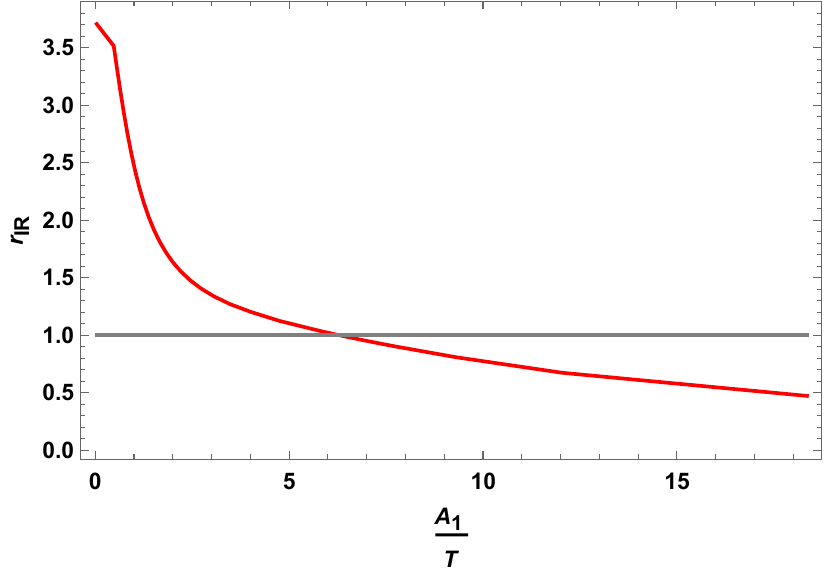}
    \end{subfigure}
    \hfill
    \begin{subfigure}{0.5\textwidth}
    \includegraphics[width=\textwidth]{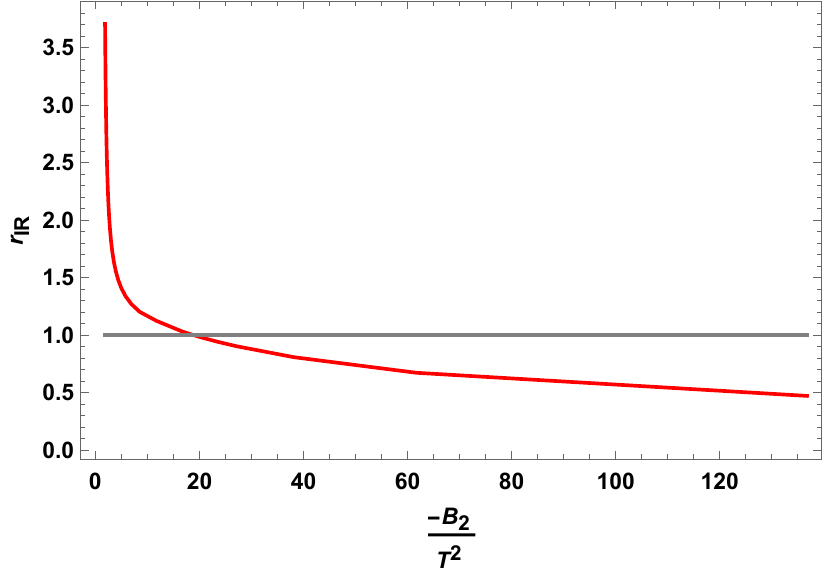}
    \end{subfigure}
    \caption{Two representative plots of $r_{\rm IR}$ with deformations for the SUGRA model. The horizontal line is showing the position of the horizon.}
    \label{rirwithdef}
\end{figure}

\begin{figure}[h]
\begin{subfigure}{0.7\textwidth}
    \centering
    \includegraphics[width=\textwidth]{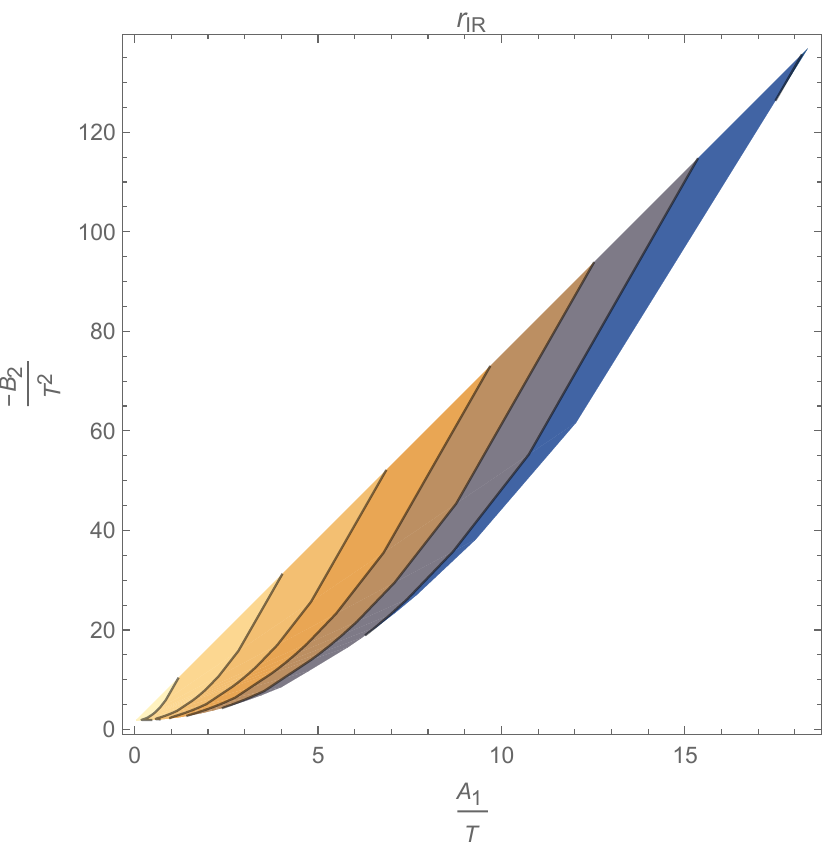}
    \end{subfigure}
    \begin{subfigure}{0.07\textwidth}
    \includegraphics[width=\textwidth]{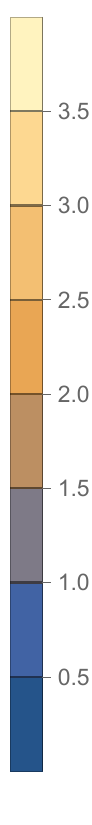}
    \end{subfigure}
    \caption{A contour plot of $r_{\rm IR}$ with the deformation parameters for the supergravity model.}
    \label{rircontour}
\end{figure}

Here equation \eqref{expression of a} reduces to 
\begin{equation*}
    a_{\beta}=\frac{\pi^2}{\Gamma(2) l_P^3} {\rm exp}\left[-\frac{3}{2}\chi(r)\right]
\end{equation*}
Figure \ref{cwithdeformationSUGRA}  shows a plot of the above $a_{\beta}$ with r, for a set of three different values of the deformation parameters.
Figure \ref{rirwithdef} is showing the position of $r_{\rm IR}$ with deformation parameters $\frac{A_1}{T}$ and $\frac{-B_2}{T^2}$. Here, the picture is different from the toy model. For very small deformation, $r_{\rm IR}$ is well inside the horizon; as the deformation increases, this point comes close to the horizon and crosses it. Though in this figures only one deformation parameter is displayed, the other deformation parameter is also changing. A more natural contour plot is shown in Figure \ref{rircontour}.
%

\section{Discussions}

In this article, we considered the effect of relevant deformation to a thermo-field double state of a CFT. In particular, we considered the TFD state of ${\cal N}=4$ SYM theory, perturbed by a mass deformation for one of the complex scalars. This deformation, usually, triggers an RG-flow that ends in a non-trivial IR fixed point, specifically the ${\cal N}=1$ SYM. For the TFD-state, however, this IR fixed point becomes inaccessible since it always remains inside the event horizon, even in the extreme limit of vanishing temperature. We have argued, on general grounds, that a TFD-state will always hide such an IR fixed point inside the horizon and within the supergravity approximation, one cannot push this IR fixed point out through the horizon.

In the first case, the dimensionful deformation allows us to define a natural vanishing temperature limit: $\phi_0 / T^{d-\Delta} \to \infty$. However, the geometric horizon never disappears and therefore the TFD state always remains entangled. Geometrically, the IR fixed point approaches the event horizon from inside, but can never touch it. This feature appears robust, both in the toy model that we considered as well as with the supergravity potential. We have explicitly shown that placing the IR fixed point on the event horizon freezes the scalars globally and this solution cannot be reached by tuning the deformation parameter smoothly.  As we have discussed, this is a consequence of an event horizon in the (semi-)classical bulk dual description which is valid strictly at large $N$. For a CFT, a vanishing mass gap is expected irrespective of its central charge.  It will further be very interesting to construct an explicit QFT-example, away from the large-$N$ limit, in which such a mass gap becomes non-vanishing as a result of a relevant deformation.

In a special case, the IR fixed point is located exactly at the event horizon and remains frozen inside of it. This class of solutions are disconnected from the type above and their interior receives no scalar back-reaction. In this case, the IR fixed point can be reached only in the limit $\phi_0/T^{d-\Delta} \to \infty$. The presence of the horizon implies a similar feature for the mass gap. From an Wilsonian perspective, we can again think of an effective theory defined at the IR fixed point which also has the AdS-isometry, {\it i.e.}~the full conformal group.\footnote{The local AdS-Rindler structure of the event horizon preserves the symmetry of an AdS-geometry. } The corresponding RG-flow, once again, stops at the event horizon.

Note that, from a geometric perspective the vanishing mass gap argument holds for any eternal black hole background. Thus, large $N$ theories with a Holographic dual, whose thermofield double state is dual to an eternal black hole, will always have a vanishing mass gap. On the other hand, there are well-known examples in Holography where the dual theory confines and develops a mass gap. For these cases, the TFD-state is not dual to an eternal black hole, rather two disjoint copies of the confined phase. Usually, such confining IR-behaviour can be obtained by deforming a UV-CFT and performing an RG-flow. Intuitively, from our discussions above, it appears that a TFD-state in the UV-CFT, deformed by the relevant operator, will never see the IR confined system within a (semi-)classical gravitational description. It would therefore be very interesting to understand how a TFD-state in the UV CFT may flow to an IR state, away from the large $N$ limit.

In a precise sense, the RG-flow geometries also capture a notion of an ``RG-flow" of quantum information theoretic observables in this framework: For example, wormholes connecting the two sides of the eternal black hole. This is true by construction, since the Einstein-Rosen bridge receives an explicit correction from the scalar back-reaction. These Wormholes, and their modifications, have been discussed in the recent literature from various aspects: traversability\cite{Gao:2016bin}, quantum teleportation protocol\cite{Maldacena:2017axo}, quantum regenesis\cite{Gao:2018yzk}, to name a few. See {\it e.g.}~\cite{Kundu:2021nwp} for a review for a summary of some such directions. It will be very interesting to explore the flow geometries that we have constructed from these points of view. The boundary CFT observables will explicitly depend on the back-reaction $\phi_0/T^{d-\Delta}$ which has a natural RG-flow interpretation. It will further be very interesting to explore a potential universality within these flows. Specifically, for example, the consequence of a symmetry in $p_t$ between the limits $\phi_0/T^{d-\Delta} \to 0$ and $\phi_0/T^{d-\Delta} \to \infty$. We hope to address some of these issues in future.

\section{Acknowledgements}

We would like to thank Elena Caceres, Diptarka Das, Ayan K.~Patra, Sanjit Shashi for numerous helpful discussions, conversations and collaborations related to the present work. We especially thank Nikolay Bobev for comments and feedback on the draft. AK is supported by the Department of Atomic Energy, Govt.~of India and CEFIPRA grant no.~6304-3.

\appendix

\section{Calculation of conformal dimension $\Delta$ at $(0,0)$}

Value of the potential at $(0,0)$ is $V_{\rm max}=\frac{-3g^2}{4}$ and the mass matrix is given by 
\begin{equation}
    \mathcal{M}=\left.\frac{\partial^2 V}{\partial\phi_i \partial \phi_j}\right\vert_{(0,0)}=\begin{pmatrix}
-\frac{3}{4}g^2  & 0 \\
0 & -g^2 
\end{pmatrix} \ , 
\end{equation}
So, the mass squared of the fields $\phi_1$ and $\phi_3$ are given by $m_1^2=-\frac{3}{4}g^2$ and $m_3^2=-g^2$. The action for a $(d+1)$ dimensional AdS black hole background with radius $l$ is given by (see \cite{Siopsis:2010pi}):
\begin{equation*}
    \mathcal{S}=\int d^{d+1}x \sqrt{\mid{g}\mid}\left(\frac{R+d(d-1)/l^2}{2\kappa_5^2}+{\rm matter} \right) \ .
\end{equation*}
Comparing this with action \eqref{action} we get $d(d-1)/l^2=-4V$ {\it i.e.}~$l^2=\frac{4}{g^2}$. The conformal dimension of a field with mass $m$ in AdS$_{d+1}$ with radius $l$ is given by
\begin{equation*}
    \Delta_{\pm}=\frac{d}{2}\pm\sqrt{\frac{d^2}{4}+m^2l^2} \ . 
\end{equation*}
Then it is straightforward to calculate that $\Delta_{\pm}=3,1$ for $\phi_1$ and $\Delta_{\pm}=2,2$ for $\phi_3$.

\section{Impossibility of Turning off the Irrelevant Deformation}\label{Appendix:relevant}

As mentioned before, the saddle points of the potential \eqref{pot2} are given by, $(\phi_1,\phi_3)=(\pm\frac{1}{2}\log3,\frac{1}{\sqrt{6}}\log2)$. Here we only consider the first one without any loss of generality. For that point the mass matrix is given by:
\begin{equation}
    \mathcal{M}=\begin{pmatrix}
\frac{4}{3}2^{\frac{1}{3}}g^2  & \frac{4}{3\sqrt{3}}2^{\frac{5}{6}}g^2\\
\frac{4}{3\sqrt{3}}2^{\frac{5}{6}}g^2 & \frac{4}{9}2^{\frac{1}{3}}g^2
\end{pmatrix} \ .
\end{equation}
The eigenvalues are given by $m_1^2=\frac{4}{9}2^{\frac{1}{3}}(2+\sqrt{7})g^2$ and $m_2^2=\frac{4}{9}2^{\frac{1}{3}}(2-\sqrt{7})g^2$. The first one corresponds to an irrelevant deformation and the second one corresponds to relevant one. One can always set this irrelevant deformation to zero at that particular point. This makes the potential to look like that of `$m^2 \phi^2$' locally but we can not set it globally because this is inconsistent with equations of motion.

The potential \eqref{pot2} around $\Phi=(\phi_1,\phi_3)=\big( p (=\frac{1}{2}\log3),q (=\frac{1}{\sqrt{6}}\log2)\big)$ can be Taylor approximated as:
\begin{equation}\label{A1}
    \begin{split}
        V &= \Phi^T \mathcal{M} \Phi \\
    &=\begin{bmatrix}
    (\phi_1-p) & (\phi_3-q)
    \end{bmatrix} 
    \begin{bmatrix}
\frac{4}{3}2^{\frac{1}{3}}g^2  & \frac{4}{3\sqrt{3}}2^{\frac{5}{6}}g^2\\
\frac{4}{3\sqrt{3}}2^{\frac{5}{6}}g^2 & \frac{4}{9}2^{\frac{1}{3}}g^2
\end{bmatrix}  
\begin{bmatrix}
    (\phi_1-p) \\
    (\phi_3-q)
    \end{bmatrix} \ .
    \end{split}
\end{equation}

To diagonalize $\mathcal{M}$, let $\Phi=\mathbf{S}  \tilde{\Phi}$, where $\tilde{\Phi}$ is new field and $\mathbf{S}$ is some orthogonal matrix. Then \eqref{A1} can be written as:
\begin{gather*}
  \begin{split}
        V &=\tilde{\Phi}^T \mathbf{S}^T \mathcal{M} \mathbf{S} \tilde{\Phi}\\
        &=\tilde{\Phi}^T  D_{\mathcal{M}}  \tilde{\Phi} \ .
    \end{split}   
\end{gather*}
Here, $ D_{\mathcal{M}}$ is a diagonal matrix corresponding to $\mathcal{M}$. $D_{\mathcal{M}}$ and $\mathbf{S}$ are given by:
\begin{gather*}
    D_{\mathcal{M}}=  \begin{bmatrix}
m_1^2  & 0\\
0 & m_2^2
\end{bmatrix}\qquad {\rm and} \qquad \mathbf{S}=\begin{bmatrix}
    a & -b \\
    b & a
\end{bmatrix} \ .
\end{gather*}
Here $a=\sqrt{\frac{1}{14}(7+\sqrt{7})}$ and $b=\sqrt{\frac{1}{14}(7-\sqrt{7})}$. The new fields are given by:
\begin{gather*}
    \begin{split}
        \tilde{\Phi} &=\mathbf{S}^{-1} \Phi\\
        &=\begin{bmatrix}
            a & b \\
            -b & a
        \end{bmatrix}
        \begin{bmatrix}
            \phi_1-p\\
            \phi_3-q
        \end{bmatrix}\\
        &=\begin{bmatrix}
            a(\phi_1-p)+b(\phi_3-q)\\
            -b(\phi_1-p)+a(\phi_3-q)
        \end{bmatrix} = \begin{bmatrix}
            \vartheta\\
            \varphi
        \end{bmatrix} \ .
    \end{split}
\end{gather*}

To set the irrelevant deformation to zero, we have to choose $\vartheta=0$ {\it i.e.}~$\phi_3=-\frac{a}{b}(\phi_1-p)+q$. Then, $\varphi=(p-\phi_1)\left(\frac{a^2+b^2}{b}\right)$. In terms of this new field $\varphi$ the potential \eqref{pot2} is given by:
\begin{equation}\label{relevantpot}
    \begin{split}
        V_{\varphi}=&\frac{1}{8} \exp\left[-2\sqrt{\frac{2}{3}} \left(q + \frac{a \varphi}{a^2 + b^2}\right)\right] g^2 \cosh^2\left(
  p - \frac{b \varphi}{a^2 + b^2}\right)\Bigg[-3 - 
   4 \exp\left[\sqrt{6}\left(q + \frac{a \varphi}{a^2 + b^2}\right)\right]+\\
   &\cosh\left[2\left(p - \frac{b \varphi}{a^2 + b^2}\right)\right]+\exp\left[2\sqrt{6}\left(q + \frac{a \varphi}{a^2 + b^2}\right)\right]\sinh^2\left[\left(p - \frac{b \varphi}{a^2 + b^2}\right)\right]\Bigg] \ . 
    \end{split}
\end{equation}
\begin{figure}[h!]
	\begin{subfigure}{0.5\textwidth}
		\centering
		\includegraphics[width=\textwidth]{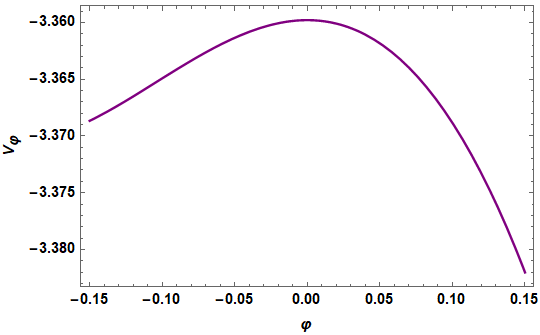}
	\end{subfigure}
	\hfill
	\begin{subfigure}{0.5\textwidth}
		\centering
		\includegraphics[width=\textwidth]{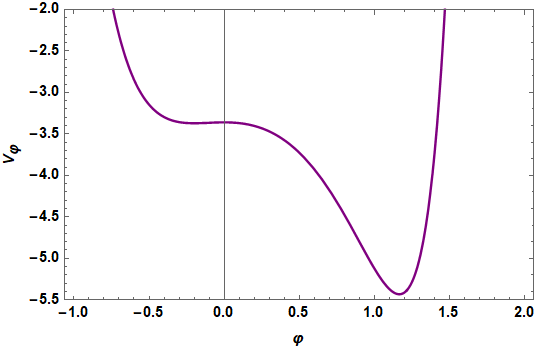}
	\end{subfigure}
	\caption{Here $V_{\varphi}$ is plotted as a function of $\varphi$. Left figure is showing the (expected) local structure of $V_{\varphi}$ near $\varphi=0$ whereas the right figure corresponds to global structure of the potential. }
	\label{fig:relevant_potential}
\end{figure}
The Potential \eqref{relevantpot} has local maxima at $\varphi=0$ with other two minima as shown in Figure \ref{fig:relevant_potential}. Although deformation at $\varphi=0$ is relevant (we set it), deformations at the minima are irrelevant. So we get a local flow from AdS to Kasner cosmology for small deformation around $\varphi=0$. Here a local flow is shown below in Figure \ref{fig:relevant_flow}.
\begin{figure}[h!]
	\begin{subfigure}{0.5\textwidth}
		\centering
		\includegraphics[width=\textwidth]{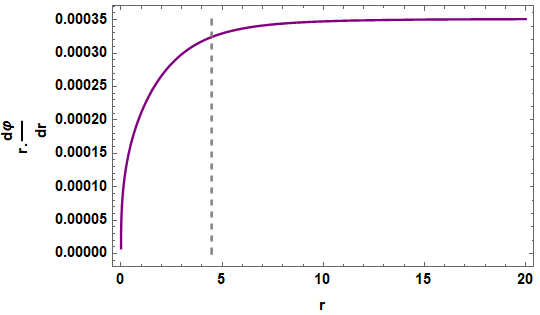}
	\end{subfigure}
	\hfill
	\begin{subfigure}{0.5\textwidth}
		\centering
		\includegraphics[width=\textwidth]{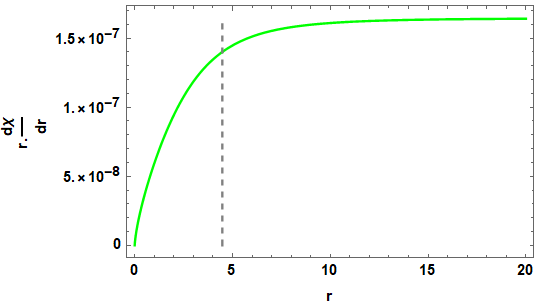}
	\end{subfigure}
	\begin{subfigure}{0.5\textwidth}
	    \centering
	    \includegraphics[width=\textwidth]{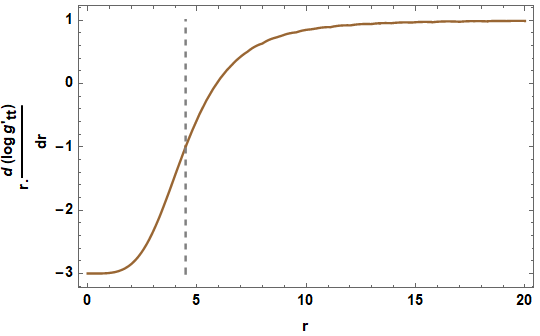}
	\end{subfigure}
	\caption{Here $V_{\varphi}$ is plotted as a function of $\varphi$. Left figure is showing the (expected) local structure of $V_{\varphi}$ near $\varphi=0$ whereas the right figure corresponds to global structure of the potential. }
	\label{fig:relevant_flow}
\end{figure}
%

\section{Geodesic Length Calculations}\label{Appendix:Integration}

In this section we discuss, in detail, the integrations appeared in \eqref{boundary time} and \eqref{geodesic length}. As mentioned before, these integrations are somewhat tricky to solve analytically because we only know the form of $f(r)$ and $\chi(r)$ towards the boundary and the singularity. As the process is almost same for \eqref{boundary time} and \eqref{geodesic length}, here we only consider the second one which is given by (ignore the counter term for the time being):
\begin{equation}\label{length1}
   L= \frac{2}{|E|}\int_{r_c}^{r_\ast} \frac{e^{\frac{-\chi}{2}}}{r^2\sqrt{1+\frac{f e^{-\chi}}{(r E)^2}}} \,dr \ . 
\end{equation}
As we are interested in large $E$ limit we can expand the integrand in small $\frac{1}{E}$ which gives the basic structure of \eqref{length1} as $\frac{L_1}{E}+\frac{L_3}{E^3}+\frac{L_5}{E^5}+\cdots$ upto some non analytic terms which come from end points.

To get the explicit result that comes from the lower end point $r_c$, substitute near boundary expansion \eqref{boundary expansion} for $f$ anf $\chi$ and substitute $x=r |E|$ in the integration. Then we expand the integrand in small $\frac{1}{E}$, integrate it term by term, and then substitute $x=r_c |E|$. Finally, we take limit $r_c\rightarrow0$ which can be taken in two ways:
\begin{enumerate}
    \item  $r_c\rightarrow0$ such that $r_c |E|\rightarrow0$. In Mathematica  this can be done by expanding the result in small $r_c$ and ignoring higher order term in $r_c$ (in doing so Mathematica treats $E$ as a fixed number).
    \\
 \item $r_c\rightarrow0$ such that $r_c|E|$ is finite. For this, expand the result in small $\frac{1}{E}$ which generates some terms like $\frac{A_1}{r_cE},\frac{A_3}{(r_c E)^3},\cdots$ for some constants $A_1,A_2,\cdots$, but we don't write them explicitly because $\frac{1}{E},\frac{1}{E^3},\cdots$ contributions also come from the whole integration and this both contributions are written in the form $\frac{l_1}{E},\frac{l_3}{E^3},\cdots$ in expression \eqref{geodesic length final}.
\end{enumerate}
 The result is:
  \begin{equation*}
       -\Bigg( -L_0+\frac{4}{g l}\log r_c-\frac{4}{g l}\log\Big(\frac{2}{E}\Big)-\frac{L_2}{E^2}-\frac{L_4}{E^4}-L_5\frac{\log E}{E^4}-L_6\frac{(\log E)^2}{E^4}+\cdots \Bigg) \ .
    \end{equation*}
As this contribution is coming from the lower end point of the integration there is an overall negative sign in the above expression. Here:
\begin{equation*}
    \begin{split}
        L_0= & \frac{8}{g l}\log (gl)+\frac{4}{gl}\log (2gl), \hspace{10 pt} L_2=\frac{gl}{3}A_1^2 \ , \\
        L_4= & -gl F_2-(gl)^3\Big[\frac{5}{108}A_1^4+\frac{2}{3}A_1A_2+\frac{2}{9}\Big(2B_1^2+2B_1B_2+B_2^2\Big)\Big]+\\
        &\frac{(gl)^3}{81}\Big[\big(B_2^2(41+6\pi^2)-30B_1B_2\big)+6\log (gl)\big(6B_1B_2-7B_2^2+6B_2^2\log (gl)\big)\Big] \ , \\
        L_5 = & \frac{2 (gl)^3}{9}\Big[\frac{2}{3}B_2^2\big(5-6\log (gl)\big)+2B_1B_2+B_2^2\Big], \hspace{10 pt} L_6=-\frac{4 B_2^2(gl)^3}{9} \ . 
        \end{split}
\end{equation*}
To integrate near the upper end point we substitute the expressions of $f$ and $\chi$ from \eqref{KasnerEq1} and expand the integrand in small $\frac{1}{E}$ and integrate term by term. Subsequently, we set $r\rightarrow r_\ast$ which is given by expression \eqref{turning point}. The result is:
\begin{equation*}
    \frac{1}{E^{\frac{c+8}{4-c}}}\Bigg[\frac{e^{\frac{\chi_1}{2p_t}}}{f_0^{\frac{1+p_t}{2P_t}}}\Big[-\frac{4}{c+2}-\frac{1}{c-1}+\frac{1}{2(2-c)}+\cdots\Big] \Bigg] \ ,
\end{equation*}
where $p_t=\frac{c-4}{c+8}$. Finally, combining this two and subtracting the diverging term $-\frac{4}{g} \log r_c$, one obtainsthe answer for regulated geodesic length \eqref{geodesic length final}.

\section{Detailed expressions of coefficients }

Here we simply collect some explicit formulae which are relevant for the discussion in section 6, for the sake of completeness. 
\begin{equation*}
    \begin{split}
        T_1 =& \frac{2}{g l}, \hspace{15 pt} T_3=\frac{A_1^2}{9 g l} \ ,\\
        \\
        T_5 =& -\frac{1}{270}\Big[108 g l F_2+(g l)^3\big(5 A_1^4+72 A_1A_2+24(2B_1^2+2B_1B_2+B_2^2)\big) \Big]+\\
        & \frac{(g l)^3}{10125}\Big[B_2^2\big(1909+300\pi^2\big)+120\log (g l)\big(15B_1B_2-16B_2^2+15B_2^2\log (g R)\big)-1410B_1B_2\Big] \ ,\\
        \\
        T_6 =&\frac{4 (g l)^3}{45}\Big[(2B_1B_2+B_2^2)+\frac{1}{15}(47-60\log (g l))B_2^2\Big], \hspace{10 pt} T_7 =-\frac{8 B_2^2 (g l)^3}{45} \ ,\\
        \\
        T_8 =& -\frac{e^{\frac{\chi_1}{2p_t}}}{f_0^{\frac{1+p_t}{2P_t}}}\Big[\frac{1}{3}+\frac{1}{2+c}+\frac{3}{8(c-1)}+\cdots\Big] \ .\\
        \\
        L_0= & \frac{8}{g l}\log (g l)+\frac{4}{g l}\log (2g l), \hspace{10 pt} L_2=\frac{g l}{3}A_1^2 \ ,\\
        \\
        L_4= & -g l F_2-(g l)^3\Big[\frac{5}{108}A_1^4+\frac{2}{3}A_1A_2+\frac{2}{9}\Big(2B_1^2+2B_1B_2+B_2^2\Big)\Big]+\\
        &\frac{(g l)^3}{81}\Big[\big(B_2^2(41+6\pi^2)-30B_1B_2\big)+6\log (g l)\big(6B_1B_2-7B_2^2+6B_2^2\log (g l)\big)\Big] \ ,\\
        \\
        L_5 = & \frac{2 (g l)^3}{9}\Big[\frac{2}{3}B_2^2\big(5-6\log (g l)\big)+2B_1B_2+B_2^2\Big], \hspace{10 pt} L_6=-\frac{4 B_2^2 (g l)^3}{9} \ ,\\
        \\
        L_7 = & \frac{e^{\frac{\chi_1}{2p_t}}}{f_0^{\frac{1+p_t}{2P_t}}}\Big[-\frac{4}{c+2}-\frac{1}{c-1}+\frac{1}{2(2-c)}+\cdots\Big]  \ . 
    \end{split}
\end{equation*}
Here $l^2=\frac{4}{g^2}$, $c$ is given by equation \eqref{EOMsingularity} and $p_t=\frac{c-4}{c+8}$.

\bibliography{bibliography1.bib}
\bibliographystyle{JHEP.bst}

\end{document}